\documentclass[preprint,prd,aps,tighten,nofootinbib,amssymb]{revtex4}
\usepackage{color}
\input{colordvi.tex}
\usepackage{amsmath}
\usepackage[dvips]{graphicx}
\usepackage{subfigure}
\newcommand{\beq}{\begin{equation}}
\newcommand{\eeq}{\end{equation}}
\newcommand{\beqa}{\begin{eqnarray}}
\newcommand{\eeqa}{\end{eqnarray}}

\newcommand{\CO}{{\cal O}}

\begin{document}
\widetext
\draft

\begin{flushright}
KUNS-2426 \\
DESY 12-218
\end{flushright}

\vspace{1cm}

\title{Cosmological constraints on spontaneous R-symmetry breaking models}

\author{Yuta Hamada}
\affiliation{Department of Physics, Kyoto University, Kyoto 606-8502, Japan}

\author{Kohei Kamada}%
\affiliation{Deutsches Elektronen-Synchrotron DESY, Notkestra\ss e 85, D-22607 Hamburg, Germany}

\author{Tatsuo Kobayashi}
\affiliation{Department of Physics, Kyoto University, Kyoto 606-8502, Japan}

\author{Yutaka Ookouchi}
\affiliation{The Hakubi Center for Advanced Research \& Department of Physics, Kyoto University, Kyoto 606-8302, Japan}

\date{\today}


\begin{abstract}
We study general constraints on spontaneous R-symmetry breaking models coming from the 
cosmological effects of the pseudo Nambu-Goldstone bosons, 
R-axions. 
They are substantially produced in the early Universe and may cause several cosmological problems. 
We focus on relatively long-lived R-axions and find that in 
a wide range of parameter space, models are severely constrained. 
In particular, R-axions with mass less than 1 MeV are generally ruled out for relatively high reheating temperature, 
$T_R>10$ GeV. 

\end{abstract}

\maketitle

\section{Introduction}

Supersymmetry (SUSY) has been considered to be the strongest candidate of the physics beyond the 
standard model (BSM). Although the recent data from the Large Hadron Collider (LHC) has not shown 
any evidence for SUSY but discovered a Standard Model Higgs-like particle with a 
mass of around 125 GeV \cite{LHChiggs}, it still remains a strong candidate of BSM because 
it suggests the gauge coupling unification, it guarantees proton stability sufficiently, and it provides a
reasonable dark matter candidate. 
Moreover, in string theories, which are the most powerful candidates of the quantum theory of gravity, 
it plays a crucial role for consistency and must be broken at a scale between the electroweak scale and 
the Planck scale.
Therefore, it is important to investigate SUSY-breaking models in the light of LHC data \cite{SUSYiloHiggs}. 

R-symmetry, which is a specific symmetry of supersymmetric models, is a key ingredient for SUSY 
breaking and its application to model building. Recent drastic progresses on SUSY breaking by exploiting a metastable state (see \cite{rev1,rev2,rev3} for reviews and references therein) gives us a better understanding of the role of the R-symmetry in realistic model building \cite{Kitano1,Kitano2,Intriligator:2007py,KOO,Abe:2007ax,R1,R2,Kang:2012fn,Ferretti:2007ec,Cho:2007yn,Abel:2007jx,Aldrovandi:2008sc,Carpenter:2008wi,Giveon:2008ne,Higaki:2011bz}. Nelson-Seiberg's argument \cite{Nelson:1993nf} beautifully demonstrates a connection between metastability and R-symmetry in the context of generalized Wess-Zumino models with a generic superpotential. If R-symmetry is preserved, there is no SUSY vacuum in a finite distance in field space. On the other hand, if a gaugino mass has Majorana mass, R-symmetry has to be broken to generate the gaugino mass. Thus, there is a tension between stability of vacuum and generating gaugino mass. A simple solution to this problem is to introduce an approximate R-symmetry. 

One of the interesting ways to break R-symmetry is spontaneous breaking. In Ref.~\cite{Shih:2007av}, D. Shih revealed a quite fascinating condition for spontaneous R-symmetry breaking in the context of generalized O'Raifeartaigh models: For R-symmetry breaking, there must be a field with R-charge different from $0$ or $2$. Such models were applied to gauge mediation \cite{Extra} and some classes of the models successfully generated large gaugino masses. According to the general argument by Komargodski and Shih \cite{KS}, large gaugino mass is related to a tachyonic direction at a point in pseudo moduli space toward the messenger direction. In the R-symmetric model, such tachyonic direction exists at the origin of the pseudo moduli space.

When the spontaneous breaking of $U(1)_R$ symmetry occurs, cosmic R-strings are formed by the Kibble-Zurek mechanism \cite{Kibble,Zurek}. Plugging the structure of the pseudo-moduli space mentioned above and R-string forming, we will meet a quite dangerous possibility. It is known as a ``roll-over'' process of vacuum through inhomogeneous energy distribution by an impurity such as a cosmic string \cite{Pole,Hosotani}. In the core of the R-string, 
the system can easily slide down to the lower vacuum via the tachyonic direction at the origin and form a sort of ``R-tube'' in which the core sits in the lower energy vacuum. Thus, if the tube is unstable, by rapid expansion of the radius, the universe can be filled by the unwanted SUSY vacuum. As discussed in Ref.~\cite{OurpaperI} this gives a constrain for model building. However, as emphasized in Ref.~\cite{NakaiOokouchi,AzeyanagiKobayashi},  when a $D$-term contribution is not negligible, it can lift the tachyonic direction and stabilize the pseudo-moduli space. In such models, the roll-over process does not occur. Also, when the amplitude of (tachyonic) messenger mass at the origin is sufficiently smaller than that of R-symmetry breaking field, the vacuum selection is successfully realized. 
As we will see, R-strings are unstable due to the explicit R-symmetry breaking term in the superpotential and hence
the roll-over process can be circumvented if the life-time due to the explicit R-symmetry breaking is shorter than 
that for the roll-over process.  
In this paper, we assume such an early stage scenario and study general cosmological constraints for the models. 
In this sense, the results shown in the present paper is complementary to the ones studied in Ref.~\cite{OurpaperI}.

In spontaneous R-symmetry breaking models, there exists a pseudo Nambu-Goldstone boson, called R-axion, as well as 
the modulus field called R-saxion. They are copiously produced in the early Universe from scattering of thermal plasma, 
coherent oscillation, R-string decay and so on, and may cause other cosmological problems. 
Note that although we commented on the importance of R-strings, there are many other sources of R-axions and 
we should take into account all the contributions at the same time. 
Model parameters on spontaneous R-symmetry breaking model can be constrained from such cosmological considerations. 
Note that, unlike the QCD-axion, R-axions receive relatively heavy mass from gravitational coupling with 
explicit R-symmetry breaking constant term in the superpotential and its lifetime can be much shorter than the cosmic age. 
Thus, we can impose not only constraints from the R-axion overclosure problem but also 
that from R-axion decay. 
In this paper, we investigate their cosmological constraints focusing on relatively long-lived parameter range. 
We show that the model parameter space is severely constrained and many parameter space of R-axion is 
ruled out from the cosmological consideration. 

This paper is organized as follows. 
In section \ref{sec2}, we explain the general feature of spontaneous R-symmetry breaking models. 
In section \ref{sec3}, we evaluate the R-axion abundance produced in the early Universe. 
Here we assume that cosmic R-string is produced in some earlier epoch. 
We list the cosmological effects induced by R-axions in section \ref{sec4}. 
We also evaluate the constraint on the parameter space from these effects. 
Section \ref{sec5} is devoted to conclusion and discussion.

\section{Spontaneous R-symmetry breaking model \label{sec2}}

In spontaneous R-symmetry breaking models, the SUSY-breaking field with a finite R-charge
acquires nonvanishing vacuum expectation value. 
The phase of the SUSY-breaking field is almost massless and identified as the Nambu-Goldstone boson. 
It acquires a small mass from explicit R-symmetry breaking term in the superpotential and called R-axions. 
In order to see its cosmological consequeces, we should first investigate their properties and interactions. 
Here, we review a simple but general R-symmetry breaking model focusing on R-axions and read off their interactions 
with several modes.

\subsection{R-symmetry breaking model}

Let us consider a simple effective superpotential for the R-charged SUSY-breaking field, $X$, integrating out the 
messenger fields, 
\begin{equation}
W_{\rm eff}=\Lambda_{\rm eff}^2 X+W_0. 
\end{equation}
Here $\Lambda_{\rm eff}$ gives the nonvanishing $F$-term for the SUSY-breaking field and 
R-symmetry breaking constant $W_0$ is introduced for the cosmological constant to vanish. 
Note that from the flat Universe condition, they are related as $\Lambda_{\rm eff}^4=3 W_0^2/M_{\rm pl}^2$ with 
$M_{\rm pl}$ being the reduced Planck mass. 
Assuming a noncanonical K\"ahler potential, $X$ can be destabilized at the origin \cite{OurpaperI}. 
Here we consider the effective potential for $X$, 
\begin{align}
V(X)&=\frac{\lambda}{4}\left(|X|^2-f_a^2\right)^2+\frac{m_a^2}{2}f_a X+{\rm h.c.} \nonumber \\
&=\frac{\lambda}{16}(\chi^2-2f_a^2)^2+\frac{m_a^2}{2\sqrt{2}} f_a \chi \cos(a/\sqrt{2}f_a),  \label{potax}
\end{align}
where we have defined $X=(\chi /\sqrt{2})e^{ia/\sqrt{2} f_a}$. 
The second term that breaks $U(1)_R$ symmetry comes from the R-symmetry breaking 
constant term in the superpotential that couples to $X$ field through the Planck
suppressed interaction in supergravity\footnote{If there are additional explicit R-symmetry 
breaking terms in the R-axion sector, 
the cross section and decay rate of R-axions are typically increased. 
Then, the constraint on the model parameters would be relaxed. However, 
introducing explicit R-symmetry breaking makes the model uncotrolable 
and hence we do not consider such extra terms here. With this assumption, 
the interactions of R-axions are represented in terms of R-axion mass $m_a$
and decay constant $f_a$ and hence the result does not depend on the detail 
of  the messenger sector or the moduli sector up to numerical factors. }. 
The R-axion mass is related to the parameters in the potential as
\begin{equation}
m_a^2=\frac{2W_0 \Lambda_{\rm eff}^2}{f_a M_{\rm pl}^2}=\frac{2\sqrt{3} m_{3/2}^2 M_{\rm pl}}{f_a}, 
\end{equation}
where $m_{3/2}=W_0/M_{\rm pl}^2$ is the gravitino mass.

Let us investigate the model further. 
We here expand $X$ around $X=f_a$ as follows, 
\begin{equation}
X=\frac{s+\sqrt{2} f_a}{\sqrt{2}} \exp (ia/\sqrt{2}f_a), 
\end{equation}
so that the fields $a$ and $s$ have canonical kinetic terms. 
Here the phase part $a$ and the radius part $s$ are identified as  R-axion and R-saxion, respectively. 
Note that the mass of R-saxion is related to the R-symmetry breaking scale as 
\begin{equation}
m_s=\sqrt{\lambda} f_a \simeq \sqrt{\frac{M_{\rm pl}}{f_a}}m_a. 
\end{equation}
In the last equality, we assumed that the K\"ahler metric is given by
\begin{equation}
g_{X{\bar X}}^{-1}\simeq 1-\frac{2{\tilde \lambda}}{f_a^2}|X|^2+\frac{\tilde \lambda}{f_a^4}|X|^4, 
\end{equation}
with $\tilde \lambda$ being a numerical constant of order of the unity, 
and $\lambda$ is related to the model parameters as $\lambda \simeq m_a^2 M_{\rm pl}/f_a^3$. 
The fermionic partner of $X$ field, ``R-axino,'' is the goldstino for the SUSY-breaking and 
absorbed in the gravitino. 
Thus, we just have to consider cosmology of gravitinos instead of R-axinos.

\subsection{Interactions of R-axions}

We now investigate interactions of the R-axion with several modes 
as well as its cross sections and decay rates. As we will see, they are useful for the cosmological 
constraints on R-axion abundance.

First of all, the R-axion to R-saxion interaction can be read off from the kinetic term of $X$, 
\begin{equation}
|\partial_\mu X|^2 \ni \frac{1}{2}\left(1+\frac{s}{\sqrt{2}f_a}\right)^2 (\partial_\mu a)^2.
\end{equation}
From this interaction, we can evaluate the decay rate of R-saxion to 2 R-axions as
\begin{equation}
\Gamma_{\rm sax}(s\rightarrow 2a) \simeq \frac{m_s^3}{64 \pi f_a^2}. 
\end{equation}

We can assign R-charges to the supersymmetric Standard Model fields  
such that the R-symmetry is consistent with all of the interactions.
After SUSY and R-symmetry are broken, 
the R-axion appears in the gaugino mass terms as well as 
the so-called B-term and A-terms.
In addition, the R-axion couplings with the gauge bosons appear through the anomaly coupling terms.
That is, the coupling between the R-axion and the photon is 
given by 
\begin{eqnarray}
\frac{C_{em}g^2_{em}}{32 \pi^2 f_a}aF_{\mu \nu} \tilde F^{\mu \nu},
\end{eqnarray}
where $F_{\mu \nu}$ is the field strength tensor of $U(1)_{em}$ 
and $C_{em}$ is the anomaly coefficient, {\it i.e.} Tr~$ U(1)_R U(1)_{em}^2$, 
which is model-dependent.
Then, the decay width of the R-axion into two photons is 
given by 
\begin{eqnarray}
\Gamma (a \rightarrow 2 \gamma) &\simeq& \frac{C_{em}^2}{16 \pi}
\left( \frac{g_{em}}{4 \pi} \right)^4 \left( \frac{m_a}{f_a} \right)^2 
m_a, \nonumber \\
 &\simeq& 6.7 \times 10^{-38} {\rm GeV} \times C_{em}^2 
\left( \frac{m_a}{1{\rm MeV}} \right)^3 
\left( \frac{10^{10}{\rm GeV}}{f_a} \right)^2.
\end{eqnarray}

Similarly, the R-axion coupling with the gluon is given by 
\begin{eqnarray}
\frac{C_gg^2_s}{32 \pi^2 f_a}aG_{\mu \nu} \tilde G^{\mu \nu},
\end{eqnarray}
where $G_{\mu \nu}$ is the SU(3) field strength tensor 
and $C_g$ is the anomaly coefficient, {\it i.e.} Tr~$ U(1)_R SU(3)^2$.
This interaction is effective in thermal production of R-axions. 
The anomaly coefficients are typically numerical factors of the order of the unity. 
It slightly changes our result but basic features do not change according to the choice of the 
coefficients. In the following, we assume $C_{em}=C_g=2$ unless we explicitly note. 

The interactions of the R-axion with the Higgs fields 
appear through the B-term.
Then, the R-axion and the Higgs fields mix each other 
in their mass terms (see for its detail Appendix \ref{app:higgs-sector}.).
The eigenstate corresponding to the low-energy R-axion $\tilde a$ includes 
the axial parts of the up and down-sector Higgs fields, 
$\xi_u$ and $\xi_d$ \cite{Goh:2008xz},
\begin{eqnarray}
\tilde a \simeq  \kappa  a + \kappa r \cos^2 \beta \sin \beta ~\xi_u
+ \kappa r \sin^2 \beta \cos \beta ~\xi_d,
\end{eqnarray}
where $r=v/(\sqrt{2}f_a)$, $v=246$ GeV, $\kappa = (1+r^2 \sin^2 2
\beta)^{-1/2}$.
Note that $a$ denotes the R-axion at high
energy beyond the electroweak symmetry breaking.
Since the coefficients of $\xi_{u,d}$ are very small, 
it is found that $\tilde a \sim a $.
Hereafter, we denote the low-energy R-axion by $a$ instead of 
$\tilde a$.
However, because of this mixing, the R-axion can couple with the quarks 
and leptons through their Yukawa couplings.
That is, the couplings of the R-axion with the up-type quarks, 
the down-type quarks and the charged leptons, $\lambda_u$, 
$\lambda_d$ and $\lambda_\ell$, are given by 
\begin{eqnarray}
\lambda_u &=& i y_u\kappa r \cos^2 \beta \sin \beta = 
i\frac{m_u}{f_a}\kappa \cos^2 \beta, \nonumber \\
\lambda_d &=& i y_d\kappa r \sin^2 \beta \cos \beta =
i\frac{m_d}{f_a} \kappa \sin^2 \beta,  \\
\lambda_\ell &=& i y_\ell \kappa  r \sin^2 \beta \cos \beta =
i\frac{m_\ell}{f_a}\kappa \sin^2 \beta, \nonumber 
\end{eqnarray}
respectively, where $y_f$ and $m_f$ with $f=u,d,\ell$ are 
their Yukawa couplings and masses. 
Through these couplings, the R-axion can decay to a pair of 
the SM fermions, if $m_a > 2m_f$.
Its decay width is given by 
\begin{eqnarray}
\Gamma(a \rightarrow f{\bar f}) = \frac{\lambda_f^2}{8 \pi} m_a 
\left( 1 - 4m_f^2/m_a^2 \right)^{1/2}.
\end{eqnarray}
For example, the decay rate into the electron pair is given by 
\begin{eqnarray}
\Gamma(a \rightarrow e^+ e^-) \simeq 1.0 \times 10^{-31} {\rm GeV} 
\times \sin^4 \beta \left( \frac{m_a}{1 {\rm MeV}} \right) 
\left( \frac{10^{10}{\rm GeV}}{f_a} \right)^2  
\left( 1 - 4m_e^2/m_a^2 \right)^{1/2}.
\end{eqnarray}
The decay rate into the $\mu$ pair is enhanced by its mass as 
$\Gamma(a \rightarrow \mu^+ \mu^-) = (m_\mu/m_e)^2 \times 
\Gamma(a \rightarrow e^+e^-) $, but such a decay occurs 
for $m_a > 2m_\mu$.

Similarly, we can compute the couplings between the 
R-axion and the neutrinos.
For the neutrinos, we consider the Weinberg operator 
in the superpotential, $y_\nu (LH_u)^2/M_R$, 
instead of the Yukawa couplings terms. 
Then, similar to the above, the coupling of 
the R-axion with neutrinos is given by 
\begin{eqnarray}
\lambda_\nu = i\frac{m_\nu}{ f_a}\kappa \cos^2 \beta.
\end{eqnarray}
Thus, the decay rate of the R-axion into the neutrino pair 
is suppressed because it is proportional to the neutrino mass 
squared, {\it i.e.} 
\begin{eqnarray}
\Gamma(a \rightarrow \nu \nu) = \left( \frac{m_\nu}{  m_e }\right)^2 \cot^4
\beta \times \Gamma(a \rightarrow ee).
\end{eqnarray}
Therefore, the branching ratio of R-axions into pair of neutrinos are small enough 
even in the case where the decay channel into electron is closed, $m_a\lesssim$ MeV. 

Note that R-axion decay associated with QCD jet production occurs when 
it is heavier than at least the proton mass, $m_a \gtrsim1$ GeV,  
which is beyond our interest. Thus,  we do not consider it here.

\begin{figure}[htbp]
\begin{center}
  \includegraphics[width=.7\textwidth]{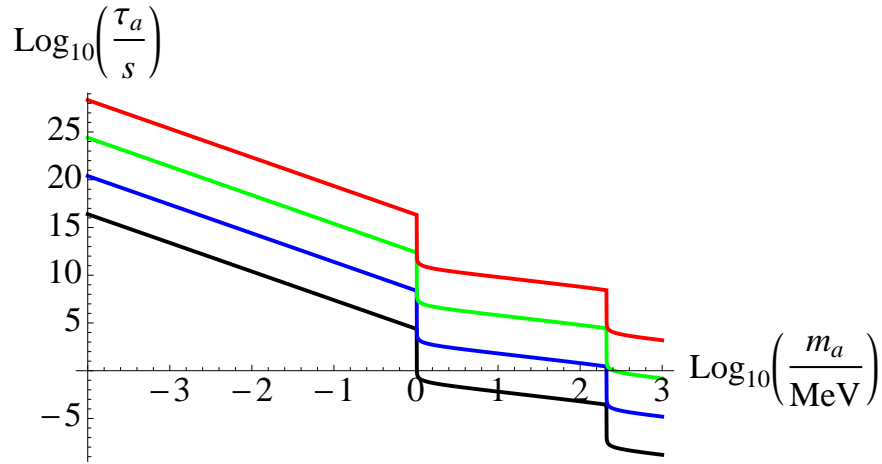}
  \caption{\sl Theoretical predictions for the R-axion lifetime  
with various values of $f_a$. 
Black, blue, green and red lines correspond to $f_a=10^{6}$GeV, $10^8$GeV, $10^{10}$GeV, and $10^{12}$GeV, respectively. Here we use $\tan \beta=30$. }
  \label{fig:life}
  \end{center}
\end{figure}

The lifetime of R-axions is given by $\tau_a \equiv \Gamma^{-1}. $ 
In Fig.~\ref{fig:life}, we show its $m_a$ dependence with each choice of $f_a=10^6, 10^8,10^{10}$ and $10^{12}$ GeV. 
We can see that the lifetime of R-axions becomes longer for smaller $m_a$ and larger $f_a$. 
We can  also see that the decay channels to electrons opens at $m_a\simeq 1$ MeV and to muons 
at $m_a \simeq 200$ MeV and the R-axion lifetime becomes shorter.

\section{R-axion production in the early Universe \label{sec3}}

Let us consider the cosmology of the spontaneous R-symmetry breaking model 
focusing on the R-axion production and evaluate the R-axion abundance. 
We consider the case where  $U(1)_R$ is restored due to some additional mass terms such as the Hubble induced mass 
or thermal mass in the early Universe\footnote{We assume that the SUSY-breaking vacuum is selected by some 
mechanism. Note that 
if the amplitude of the mass of $X$ at the origin is larger than that of messenger fields, 
the SUSY-breaking vacuum is naturally 
selected. However, this issue is beyond the scope of this study and hence we do not impose any constraints on 
the model parameters from the vacuum selection. }. 
After some epoch, $X$ field is destabilized as the additional mass term decreases and 
acquires vacuum expectation value $\chi=f_a$. 
Since the approximate $U(1)_R$ symmetry breaks spontaneously at that time, (unstable) cosmic strings are formed
by the Kibble-Zurek mechanism. 
The long cosmic strings in a Hubble volume intersect each other and generates closed string loops\footnote{
Here we do not consider the effect of the existence of SUSY vacuum  on the cosmic string structure. 
This issue will be studied elsewhere \cite{OurpaperI}. }. 
These closed string loops shrink with emitting R-axions. 
As a consequence, the cosmic string network enters the scaling regime. 
As the Hubble parameter decreases, the explicitly R-symmetry breaking term in the 
potential becomes no longer irrelevant to the dynamics of the system 
and the string network turns to the string-wall system where domain walls are attached to 
cosmic strings \cite{Sikivie:1982qv,Hiramatsu:2012gg}. 
The string-wall networks are unstable and 
annihilate when the domain wall tension becomes comparable to that of cosmic strings. 
The energy stored in the string-wall system turns to R-axion particles. 
The fate of R-axions produced from the cosmic string loops and the string-wall system as well as 
the scattering of thermal plasma and the vacuum misalignment is determined by the lifetime of R-axions, 
which, then, constrain the model parameters of spontaneous R-symmetry breaking models\footnote{
R-axions are also produced from R-saxion decay. However, as shown in Appendix \ref{app2}, 
the abundance of such R-axions are generally subdominant and hence we do not consider it here.}. 
In the following, we estimate the R-axion abundance from each source. 
We will examine the cosmological constraints in Sec. \ref{sec4}. 

\subsection{R-axion production from vacuum misalignment}
First we evaluate the energy density of the coherent oscillation of the R-axion field \cite{Dine:1982ah}. 
After the spontaneous R-symmetry breaking phase transition, the R-axion field 
acquires some initial value, $a_i$, and keeps its position after a while 
due to large Hubble friction. 
When the Hubble parameter decreases to the R-axion mass, 
\begin{equation}
H (t_{\rm osc}) =m_a, 
\end{equation}
the R-axion field starts to oscillate. 
Here the subscription ``osc'' indicates that the parameter or variable is evaluated at 
the onset of the R-axion oscillation. 
The energy density of the oscillating R-axion $\rho_{a,{\rm o}}$ is given by
\begin{equation}
\rho_{a,{\rm o}}(t_{\rm osc})=\frac{1}{2} m_a^2 a_i^2. 
\end{equation}
If the R-symmetry is broken after inflation, the initial value of R-axion 
distributes randomly from $-\sqrt{2} \pi f_a$ to $\sqrt{2}\pi f_a$ since the correlation length of R-axion 
becomes much shorter than the Hubble length at the onset of the 
R-axion oscillation. 
Therefore, we estimate the mean value of $a_i$ as
\begin{equation}
\langle a_i^2 \rangle = \frac{1}{2 \sqrt{2} \pi f_a}\int_{-\sqrt{2}\pi f_a}^{\sqrt{2}\pi f_a} a_i^2 d a_i=\frac{2\pi^2 f_a^2}{3}. 
\end{equation}

Since the energy density of R-axion oscillation decreases as $a^{-3}$, 
the quantity $\rho_a/s$ is conserved as long as there are no entropy production, where $s$ is the entropy density. 
Therefore, we characterize the axion abundance by this quantity as
\begin{equation}
\frac{\rho_{a,{\rm o}}}{s}\simeq \left\{
\begin{array}{ll}
\dfrac{15}{2g_{*s}(T_{\rm osc})}\dfrac{m_a^2 f_a^2}{T_{\rm osc}^3}, & \text{for} \quad H_{\rm osc}<H_R \\
\dfrac{\pi^2}{12} \dfrac{g_*(T_R)}{g_{*s}(T_R)} \left(\dfrac{f_a}{M_{\rm pl}}\right)^2 T_R, & \text{for} \quad H_{\rm osc}>H_R
\end{array}\right.
\end{equation}
where $g_*$ and $g_{*s}$ are (effective) relativistic degrees of freedom for energy density and entropy, respectively, 
and the subscript ``$R$'' represents that the parameter or variable is evaluated at reheating. 
Note that $T_{\rm osc}$ is given by
\begin{equation}
T_{\rm osc}=\left(\frac{90}{\pi^2 g_*(T_{\rm osc})}\right)^{1/4}m_a^{1/2}M_{\rm pl}^{1/2} \simeq 2.2 \times 10^7 {\rm GeV}\left(\frac{m_a}{1{\rm MeV}}\right)^{1/2}. 
\end{equation}
Here we assume that the scale factor increases like matter dominated era during inflaton oscillation dominated era and 
take into account the dilution until the inflaton decay or reheating when $H_{\rm osc}>H_R$.

\subsection{R-axion production from global cosmic strings}
Next we evaluate the energy density of R-axions radiated from the cosmic string loops 
\cite{Davis:1986xc,Yamaguchi:1998gx} following the discussion in 
Appendix B of Ref.~\cite{Hiramatsu:2010yu}. 
When the R-string network enters the scaling regime, 
the energy density of the long R-strings are estimated as 
\begin{equation}
\rho_{\infty} (t)=\frac{2 \pi \xi}{t^2} f_a^2 \ln \left(\frac{t/\sqrt{\xi}}{d_{\rm string}}\right). 
\end{equation}
Here the scaling parameter $\xi \simeq 0.9$ \cite{Hiramatsu:2010yu,Yamaguchi:2002zv} 
represents the mean number of strings in a Hubble volume 
and $d_{\rm string}\simeq \lambda^{-1/2} f_a^{-1}$ represents the core width of R-string. 
Note that the line energy density or the tension of R-string is given by \cite{stringreveiw}
\begin{equation}
\mu_{\rm string} \simeq 2 \pi f_a^2 \ln \left(\frac{t/\sqrt{\xi}}{d_{\rm string}}\right). 
\end{equation}
Assuming all the energy loss of long R-strings is converted into R-axion particles through 
the string loops, we obtain the evolution equations 
\begin{align}
\frac{d \rho_\infty (t)}{dt}&=-2H \rho_\infty(t)-\Gamma_{\rm em}(t),  \\
\frac{d \rho_{a,{\rm str}}(t)}{dt}&=-4 H \rho_{a,\rm str}(t)+\Gamma_{\rm em}(t), 
\end{align}
where the energy emission rate from the string loops, 
\begin{equation}
\Gamma_{\rm em}(t)=\frac{2\pi \xi f_a^2}{t^3} \times \left\{ 
\begin{array}{ll}
\left(\ln \left(\dfrac{t/\sqrt{\xi}}{d_{\rm string}}\right)-1\right), & \text{for RD} \\
\left(\dfrac{2}{3}\ln \left(\dfrac{t/\sqrt{\xi}}{d_{\rm string}}\right)-1\right).  & \text{for MD}
\end{array}\right.
\end{equation}
Here we assume that R-axion particles released from cosmic string loops are relativistic. 
Since the mean comoving momentum of radiated R-axion can be evaluated as
\begin{equation}
\frac{k_{a,{\rm str}}(t)}{R(t)}=\frac{2 \pi \epsilon}{t}, 
\end{equation}
with the constant $\epsilon \simeq 0.25$ \cite{Hiramatsu:2010yu,Yamaguchi:1998gx}, 
we can estimate the number density of radiated R-axions as
\begin{align}
n_{a,{\rm str}}(t)&=\frac{1}{R(t)^3}\int_{t_*}^t  dt^\prime \frac{R^4(t^\prime)}{k_{a,{\rm str}}(t^\prime)}\Gamma_{\rm em}(t^\prime) \notag \\
&\simeq \frac{2 \xi f_a^2}{\epsilon t} \times \left\{
\begin{array}{ll}
\left(\ln \left(\dfrac{t/\sqrt{\xi}}{d_{\rm string}}\right)-3\right), & \text{for} \quad t>t_R \\
\dfrac{1}{3}\left(\ln \left(\dfrac{t/\sqrt{\xi}}{d_{\rm string}}\right)-\dfrac{5}{2}\right).  & \text{for} \quad t<t_R
\end{array}\right.
\end{align}
Here $t_*$ is the time when the R-string network enters the scaling regime. 

When the Hubble parameter becomes comparable to the R-axion mass and R-symmetry breaking mass term 
becomes no longer irrelevant, $t=t_{\rm osc}$, string-wall system forms and R-axion emission from R-string loops stops. 
We can evaluate the resultant number density of R-axions from the R-string loops as 
\begin{equation}
n_{a,{\rm str}}(t_{\rm osc})=\frac{\xi m_a f_a^2}{\epsilon}\times \left\{
\begin{array}{ll}
4 \left(\ln\left(\dfrac{1}{2 m_a \sqrt{\xi} d_{\rm string}}\right)-3\right), & \text{for}\quad H_{\rm osc}<H_R \\
\left(\ln\left(\dfrac{2}{3 m_a \sqrt{\xi} d_{\rm string}}\right)-\dfrac{5}{2}\right).  & \text{for}\quad H_{\rm osc}>H_R
\end{array}\right.
\end{equation}
The radiated R-axions become nonrelativistic after some epoch. 
Therefore, we can approximate the R-axion energy density as $\rho_{a,{\rm str}}=m_a n_{a,{\rm str}}$ and 
the R-axion energy-to-entropy ratio as 
\begin{equation}
\frac{\rho_{a,{\rm str}}}{s}=\left\{
\begin{array}{ll}
\dfrac{90}{\pi^2} \dfrac{\xi}{ g_{*s}(T_{\rm osc})\epsilon} \dfrac{m_a^2 f_a^2}{T_{\rm osc}^3}\left(\ln\left(\dfrac{1}{2 m_a \sqrt{\xi} d_{\rm string}}\right)-3\right), & \text{for}\quad H_{\rm osc}<H_R \\
\dfrac{g_*(T_R) \xi}{4g_{*s}(T_R)\epsilon}\left(\dfrac{f_a}{M_{\rm pl}}\right)^2 T_R\left(\ln\left(\dfrac{2}{3 m_a \sqrt{\xi} d_{\rm string}}\right)-\dfrac{5}{2}\right).  & \text{for}\quad H_{\rm osc}>H_R
\end{array}\right.
\end{equation}

\subsection{R-axion production from string-wall system}

Let us evaluate the energy density of R-axions from the string-wall system annihilation 
\cite{Sikivie:1982qv,Hiramatsu:2012gg}. 
At $t\simeq t_{\rm osc}$, the explicitly R-symmetry breaking term in the potential \eqref{potax} becomes no longer 
irrelevant, and string-wall system forms. 
The surface mass density of domain walls are estimated as \cite{stringreveiw}
\begin{equation}
\sigma_{\rm wall}=16 m_a f_a^2. 
\end{equation}
When the tension of domain walls dominates that of strings, 
\begin{equation}
\sigma_{\rm wall}=\frac{\mu_{\rm string}}{t} \Leftrightarrow t \ln\left(\frac{d_{\rm string}}{t/\sqrt{\xi}}\right) = \frac{\pi}{8} m_a^{-1}, 
\end{equation}
the string-wall system annihilates. 
As following the discussion in Ref.~\cite{Hiramatsu:2012gg}, we assume that 
the energy stored in the string-wall system released to R-axion particles. 
Thus, we evaluate the number density of R-axions as 
\begin{align}
n_{a,{\rm sw}}(t)&= \frac{\rho_{\rm wall}(t_{\rm osc})+\rho_\infty(t_{\rm osc})}{\omega_a}\left(\frac{R(t_{\rm osc})}{R(t)}\right)^3 \notag \\
&=  \frac{1}{\alpha_w m_a} \left({\cal A}\frac{\sigma_{\rm wall}}{t_{\rm osc}}+\xi \frac{\mu_{\rm string}(t_{\rm osc})}{t_{\rm osc}^2}\right) \left(\frac{R(t_{\rm osc})}{R(t)}\right)^3,
\end{align}
where $\omega_a=\alpha_w m_a$ is the average energy of radiated axions and 
${\cal A} \equiv \rho_{\rm wall} t/\sigma_{\rm wall} \simeq 0.5$ \cite{Hiramatsu:2012gg} 
is the area parameter of domain walls. 
The radiated R-axions become eventually nonrelativistic and hence we can evaluate the 
energy-to-entropy ratio as
\begin{align}
\frac{\rho_{a,{\rm sw}}}{s}&=\frac{m_a n_{a,{\rm sw}}}{s}=\left\{
\begin{array}{ll}
\dfrac{180}{\pi^2 g_{*s}(T_{\rm osc})\alpha_w }\dfrac{m_a^2 f_a^2}{T_{\rm osc}^3}\left(4 {\cal A}+ \pi \xi \ln \left(\dfrac{1}{2 m_a \sqrt{\xi} d_{\rm string}}\right)\right), & \text{for}\quad H_{\rm osc}<H_R\\
\dfrac{g_*(T_R)}{4 g_{*s}(T_R)\alpha_w }\left(\dfrac{f_a}{M_{\rm pl}}\right)^2 T_R \left(24 {\cal A}+\dfrac{9 \pi}{2} \xi \ln \left(\dfrac{2}{3 m_a \sqrt{\xi}d_{\rm string}}\right)\right).  & \text{for} \quad H_{\rm osc}>H_R
\end{array}\right.
\end{align}

\ 

Noting that the logarithmic factor is evaluated as $
\ln(1/m_a \sqrt{\xi} d_{\rm string} )\simeq \ln(\sqrt{\lambda/\xi}(f_a/m_a)) =30$ for
$f_a\simeq 10^{10}$ GeV and $m_a \simeq$ 1 MeV, 
hereafter we approximate the R-axion abundance from R-axion dynamics, {\it i.e.}, 
the coherent oscillation, the decay of cosmic string loops, and the decay of the string-wall system, 
\begin{align}
\frac{\rho_{a,{\rm dyn}}}{s}&\equiv \frac{\rho_{a,{\rm o}}+\rho_{a,{\rm str}}+\rho_{a,{\rm sw}}}{s}=\left\{
\begin{array}{ll}
K_1 \dfrac{m_a^2 f_a^2}{T_{\rm osc}^3}, & \text{for}\quad H_{\rm osc}<H_R\\
K_2 \left(\dfrac{f_a}{M_{\rm pl}}\right)^2 T_R,  & \text{for} \quad H_{\rm osc}>H_R
\end{array}\right. \notag \\
& \simeq \left\{
\begin{array}{ll}
 9.4 \times 10^{-9} {\rm GeV} K_1 \left(\dfrac{m_a}{1{\rm MeV}}\right)^{1/2} \left(\dfrac{f_a}{10^{10}{\rm GeV}}\right)^2, &\text{for}\quad H_{\rm osc}<H_R\\
 1.7 \times 10^{-11}{\rm GeV} K_2 \left(\dfrac{f_a}{10^{10}{\rm GeV}}\right)^2 \left(\dfrac{T_R}{10^6{\rm GeV}}\right),  & \text{for} \quad H_{\rm osc}>H_R
\end{array}\right.
\label{dynamics}
\end{align}
where $K_1 \simeq \CO(1)$ and $K_2\simeq \CO(10)$ are numerical parameters.

\subsection{R-axion production from thermal bath}

We have estimated the abundance of R-axions generated from their dynamics. 
We should also take into account that generated from other sources. 
Here we evaluate the R-axion abundance from thermal bath. 
The R-axion abundance from R-saxion decay is discussed in Appendix \ref{app2} 
and is generally negligible. 

R-axions are produced in the thermal plasma from (mainly) gluon scattering, $gg\rightarrow ag$. 
Since the gluon-axion interaction comes from the anomaly term, 
\begin{equation}
{\cal L}=\frac{C_g g_s^2}{32\pi^2 f_a}a G^b_{\mu\nu}{\tilde G}^{b\mu\nu}, 
\end{equation}
with $C_g$ being the model dependent anomalous coefficient and $g_s$ being the strong gauge coupling, 
the R-axion abundance is calculated as \cite{Masso:2002np,Sikivie:2006ni,Graf:2010tv}, 
\begin{equation}
\frac{\rho_{a,{\rm th}}}{s} \simeq 2.0 \times 10^{-6} {\rm GeV} g_s^6 C_g^2\left(\frac{m_a}{1 {\rm MeV}}\right)\left(\frac{10^{10} {\rm GeV}}{f_a}\right)^2\left(\frac{T_R}{10^6{\rm GeV}}\right). 
\label{th1}
\end{equation}
Note that R-axions are thermalized once if the reheating temperature is high enough, 
\begin{equation}
T_R>T_D\simeq 10^6{\rm GeV} g_s^{-6} C_g^{-2} \left(\frac{f_a}{10^{10}{\rm GeV}}\right)^2, 
\end{equation}
where $T_D$ is R-axion decoupling temperature. In this case, the R-axion abundance is evaluated as
\begin{equation}
\frac{\rho_{a,{\rm th}}}{s} \simeq 2.6 \times 10^{-6} {\rm GeV} \left(\frac{m_a}{1{\rm MeV}}\right). 
\label{th2}
\end{equation}
Note that R-axion is produced thermally only if $T_R\gtrsim m_a$ is satisfied.

\begin{figure}[t]
\begin{center}
	\hfill
  \includegraphics[width=.44\textwidth]{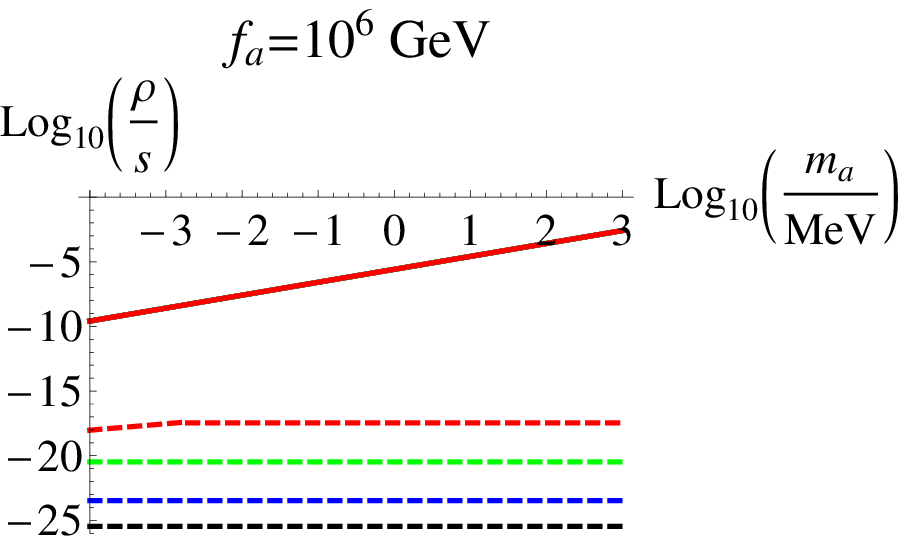}
	\hfill
  \includegraphics[width=.44\textwidth]{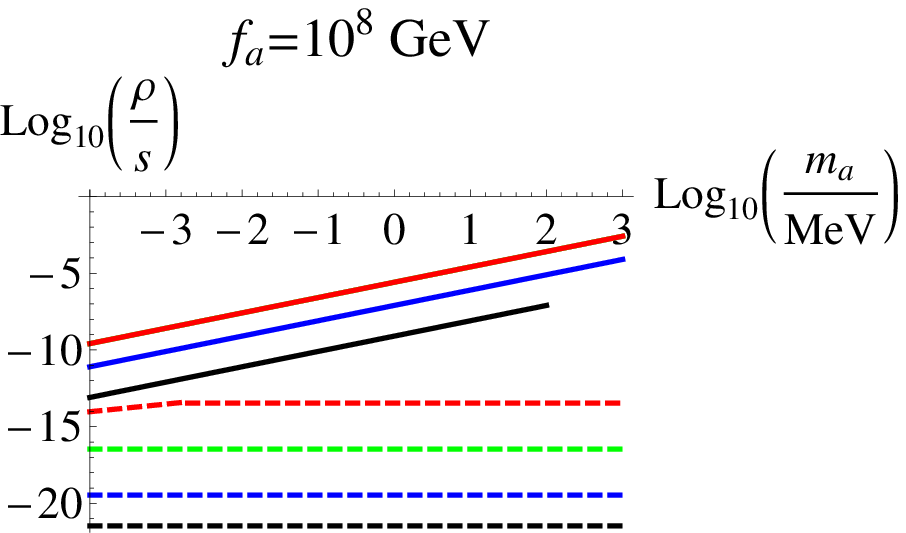}
	\hfill\mbox{}
\\
	\hfill
  \includegraphics[width=.44\textwidth]{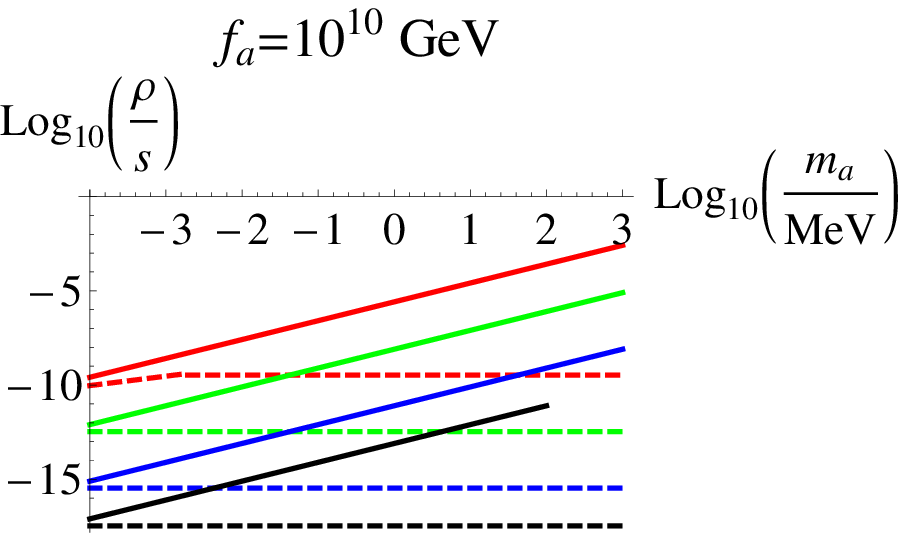}
	\hfill
  \includegraphics[width=.44\textwidth]{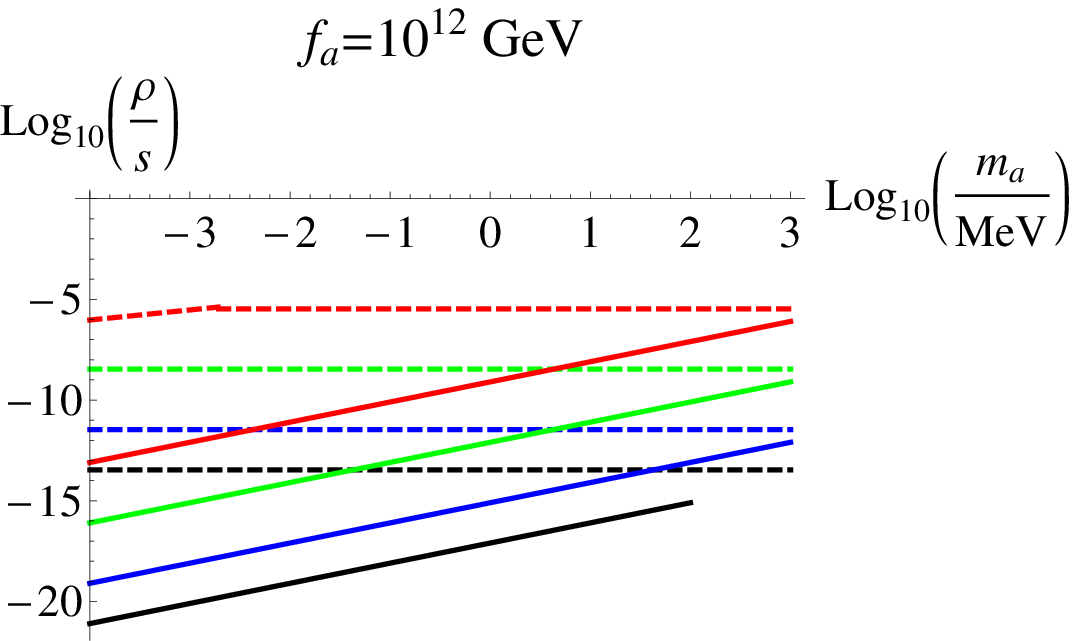}
	\hfill\mbox{}
  \caption{\sl Theoretical predictions for the R-axion to entropy ratio with $f_a=10^6$GeV, $10^8$GeV, $10^{10}$GeV, and $10^{12}$GeV. 
The solid lines represent contribution from the thermal production (Eqs.~\eqref{th1} and \eqref{th2}) 
and the dashed ones represent the R-axion dynamics (Eq.~\eqref{dynamics}) with $K_1=1$, $K_2=20$. 
Black, blue, green and red lines correspond to $T_R=10^{-2}$GeV, $1$GeV, $10^3$GeV, and $10^6$GeV, respectively. }
  \label{fig:string vs thermal}
  \end{center}
\end{figure}

As a result, the total R-axion abundance in the early Universe is evaluated by the sum of these contributions and given by
\begin{equation}
\frac{\rho_a}{s}=\dfrac{\rho_{a,{\rm dyn}}}{s}+\dfrac{\rho_{a,{\rm th}}}{s}. \label{totrax}
\end{equation}
In Fig.~\ref{fig:string vs thermal}, we show the theoretical predictions for the R-axion to entropy ratio with $f_a=10^6$GeV, $10^8$GeV, $10^{10}$GeV, 
and $10^{12}$GeV. Here the solid lines represent contribution from the thermal production (Eqs.~\eqref{th1} and \eqref{th2})
and dashed ones represent the R-axion dynamics (Eq.~\eqref{dynamics}) with $K_1=1$, $K_2=20$, respectively. Black, blue, green and red lines correspond to $T_R=10^{-2}$GeV, $1$GeV, $10^3$GeV, and $10^6$GeV, respectively. In the case of $T_R>T_D$, R-axion abundance from thermal production is independent of $T_R$. We can see that the contribution from the R-string dynamics and other R-axion dynamics generally dominates for $f_a\gtrsim10^{12}$GeV and smaller $m_a$. 
Vice versa, thermal R-axion production dominates for $f_a\lesssim10^{12}$GeV. 
Anyway, we will compare the total R-axion abundance expressed in Eq. \eqref{totrax}, 
including those from R-string dynamics and thermally produced ones, to the 
cosmological constraints discussed in the next section and will give the constraints on the 
model parameters.

\section{Cosmological constraints from R-axion \label{sec4}}

Now we consider the generic constraints of the R-symmetry breaking model from cosmology. 
One may think that the model with long-lived R-axions is safe 
if they never dominate the energy density of the Universe or
R-axions are responsible for the dark matter in the present Universe. 
However, even if they are subdominant component of the Universe, their (partial) decay
is constrained by several cosmic/astrophysical observations depending on their abundance \cite{Kawasaki:2007mk}. 
Since we have evaluated the R-axion abundance and its lifetime, 
we can constrain the model from various observations. 
As we will see, strong constraints for the model parameters are imposed. 

\subsection{Cosmological constraints on axion abundance}

Let us see the various constraints of R-axion abundance from cosmology and astrophysical observations. 
We will compare all these constraints on the R-axion abundance to that evaluated in the 
previous section, especially in Eq. \eqref{totrax} and translate them in the constraints on the 
R-axion model parameters in the next subsection.
Note that our cosmological constraints are basically 
irrelevant to what is the dominant source of R-axions, 
but relevant to the total R-axion abundance in Eq.(41).

\subsubsection{Big Bang Nucleosynthesis}
The R-axion decay into photon or electron (radiative decay) after the Big Bang Nucleosynthesis (BBN) epoch 
may break the light elements and the R-axion abundance is constrained \cite{Kawasaki:2004qu}. 
The radiative decay of R-axion causes photo-dissociation process of light elements and 
changes the light elements abundance. 
We can read off the constraint on the R-axion abundance at its decay from Ref.~\cite{Kawasaki:2004qu} as
\begin{equation}
B_{\rm r} \frac{\rho_a}{s} \lesssim
 \left\{
\begin{array}{ll}
10^{-8} {\rm GeV} \left(\dfrac{\tau_a}{10^4{\rm s}}\right)^{-2}, & \text{for}\quad 10^4 {\rm s}<\tau_a<10^7{\rm s} \\
10^{-14}{\rm GeV}, &\text{for}\quad 10^7{\rm s}<\tau_a<10^{12}{\rm s}
\end{array} \right.
\end{equation}
where $B_{\rm r}$ is the radiative branching ratio\footnote{
If the R-axion mass is heavy enough, $m_a>2$ GeV, the hadronic decay channel opens. 
In this case, more stringent constraints are imposed \cite{Kawasaki:2004qu}.}. 
Note that this effect is negligible if the energy of the injected photons is so small that
they cannot destroy the light elements. 
Thus, we here impose a condition for this constraint to be effective, 
\begin{equation}
m_a\gtrsim 4.5 {\rm MeV}, 
\end{equation}
which corresponds to the threshold energy for the deuteron destruction process, $D+\gamma \rightarrow n+p$.

\subsubsection{Cosmic microwave background distortion}
The radiative decay of R-axion before the recombination may distort the blackbody spectrum of 
CMB. After  the double-Compton scattering freezes out at $t\simeq 10^6$ s, 
energy injections generate nonzero chemical potential $\mu$ of the CMB spectrum, 
which imposes the constraint from the blackbody spectrum distortion of CMB.  
Energy injections after $t\simeq 10^9$ s, when the Compton scattering is no longer in thermal equilibrium, 
thermalize electron, which causes the Sunyaev-Zel'dovich (SZ) effect. 
Since the SZ effect is constrained by the Compton $y$-parameter, 
we can impose a constraint on the R-axion abundance. 

The COBE FIRAS measurement \cite{Fixsen:1996nj} constrains the CMB distortion as
\begin{equation}
|\mu| \lesssim 9\times 10^{-5}, \quad y \lesssim 1.2 \times 10^{-5}. \label{wai}
\end{equation}
Since the injected energy is related to these parameters as \cite{Ellis:1990nb,Hu:1993gc}
\begin{align}
\frac{\delta \rho_\gamma}{\rho_\gamma} &\sim 0.714 \mu, \quad \text{for}\quad 10^6{\rm s}<\tau_a<10^9{\rm s} \\
\frac{\delta \rho_\gamma}{\rho_\gamma} & \sim 4y, \quad \text{for}\quad 10^9{\rm s}<\tau_a<10^{13}{\rm s}
\end{align}
the constraints on the R-axion abundance is given by
\begin{equation}
B_{\rm r}\frac{\rho_a}{s} \lesssim 10^{-12}{\rm GeV} \left(\frac{10^9{\rm s}}{\tau_a}\right)^{1/2} \quad \text{for}\quad 10^6{\rm s}<\tau_a<10^{13}{\rm s} 
\end{equation}
depending on its life time.  
Note that $\mu$ and $y$-parameters impose almost the same constraint on the R-axion abundance at its decay. 

\subsubsection{Diffuse X-ray and $\gamma$-ray background}
The R-axion decay to photons after recombination, $t>10^{13}$ s, may be constrained from the 
diffuse X-ray and $\gamma$-ray background observation. 
Photons with energy $1 {\rm keV}<E_\gamma<1{\rm TeV}$ rarely scatter with the CMB photons and 
intergalactic medium. 
Therefore, the photons produced from the R-axion decay in the ``transparency window'' \cite{Chen:2003gz}, 
\begin{equation}
\tau_a \gtrsim \left\{
\begin{array}{ll}
10^{19} \ {\rm s}\left(\dfrac{m_a}{1 \ {\rm keV}}\right)^{-2}, & \text{for} \quad 1 \ {\rm keV}\lesssim m_a \lesssim 100 \  {\rm keV} \\
4 \times 10^{14} \ {\rm s}, &  \text{for}\quad 100  \ {\rm keV} \lesssim m_a \lesssim 2.5 \  {\rm MeV} \\
10^{13} \ {\rm s} \left(\dfrac{m_a}{100  \ {\rm MeV}} \right)^{-1}, & \text{for} \quad 2.5  \ {\rm MeV} \lesssim m_a \lesssim 100 \ {\rm MeV} \\
10^{13} {\rm s}, & \text{for} \quad 100  \ {\rm MeV} \lesssim m_a \lesssim 10  \ {\rm GeV}
\end{array}\right.
\end{equation}
propagate through the Universe and can be detected as diffuse background. 

The flux of the extragalactic diffuse photons is roughly given by
\begin{equation}
F^{\rm obs}_\gamma (E)/ {\rm cm}^{-2} {\rm s}^{-1}{\rm str}^{-1}  \simeq  \left\{
\begin{array} {ll}
2 \times \left(\dfrac{E}{\rm keV}\right)^{-0.4}, &0.25 {\rm keV}<E<10 {\rm keV} \\
\dfrac{3}{(E/30{\rm keV})^{0.3}+(E/30{\rm keV})^{1.9}}, & 10 {\rm keV}<E<800 {\rm keV} \\
5.0\times 10^{-3}\left(\dfrac{E}{\rm MeV}\right)^{-1.4}, & 800 {\rm keV}<E<30 {\rm MeV} \\
1.7 \times 10^{-5} \left(\dfrac{E}{100{\rm MeV}}\right)^{-1.1}, & 30{\rm MeV}<E<100{\rm MeV} \\
1.45 \times10^{-5} \left(\dfrac{E}{100{\rm MeV}}\right)^{-1.4}. & 100{\rm MeV}<E<100{\rm GeV} 
\end{array} \right.
\end{equation}
Here we applied the observational results of  ASCA \cite{ASCA} for 0.25-10 keV, HEAO  \cite{HEAO} 
for 25 keV- 800 keV, COMPTEL \cite{1998PhDT3K} for 800 keV-30 MeV, EGRET \cite{Sreekumar:1997un}
for 30 - 100 MeV, and Fermi \cite{Abdo:2010nz} for 100 MeV-100 GeV. 
Note that we have taken into account the resolved source of diffuse X-ray background 
\cite{Worsley:2004tc,Hickox:2005dz} and used the fitting formula derived in Ref.~ \cite{CXBfit}. 

The flux of photons produced from the R-axion decay can be approximated as
\begin{equation}
F_\gamma (E)\simeq B_\gamma\times \left\{
\begin{array}{ll}
 \dfrac{n_{a,0}}{2 \pi \tau_a H_0}, & \text{for} \quad \tau_a>t_0 \\
\dfrac{3 n_{a,{\rm dec} }}{4\pi} \dfrac{s_0}{s_{\rm dec}}, & \text{for} \quad \tau_a<t_0 
\end{array}\right.
\end{equation}
where the subscriptions ``0'' and ``dec'' indicate that the parameter or variable is 
evaluated at the present and the R-axion decay time, respectively, and $B_\gamma$ 
is the branching ratio to photons. 
Note that the energy of photons should be evaluated at $E=m_a/2$ for $\tau_a>t_0$ and 
$E= (3 H_0 \tau_a \sqrt{\Omega_{\rm m}}/2)^{2/3} (m_a/2)$ for $\tau_a<t_0$, 
taking into account of the redshift of the photons. 
Then, the abundance of the R-axions are constrained from the constraint $F_\gamma(E)<F_\gamma^{\rm obs}$ as\footnote
{Most of diffuse extragalactic X-ray and $\gamma$-ray background can be explained by astrophysical 
sources such as blazers. However, here we use the conservative constraint, though the flux of the unresolved 
extragalactic defuse background photon would be much more smaller when we assume some cosmological models 
of the evolution of galaxies. } 
\begin{equation}
B_\gamma \frac{\rho_a}{s} \lesssim \left\{
\begin{array}{ll}
2.4 h \times 10^{-18} {\rm GeV} \left(\dfrac{m_a}{1{\rm MeV}}\right) \left(\dfrac{\tau_a}{10^{18}{\rm s}}\right) \left(\dfrac{F_\gamma^{\rm obs}(m_a/2)}{10^{-2}{\rm cm}^{-2} {\rm s}^{-2}}\right), & \text{for} \quad \tau_a>t_0 \\ 
4.8 \times 10^{-19} {\rm GeV} \left(\dfrac{m_a}{1{\rm MeV}}\right)\left(\dfrac{F_\gamma^{\rm obs}(E)}{10^{-2}{\rm cm}^{-2} {\rm s}^{-2}}\right),  & \text{for} \quad \tau_a<t_0 
\end{array}\right. 
\end{equation}
where $h\equiv H_0/(100 \ {\rm km} \ {\rm sec}^{-1}{\rm Mpc}^{-1})$ and $H_0$ is the present Hubble parameter.

\subsubsection{Reionization}
The radiative decay of R-axion after recombination is also constrained from reionization. 
If the energy of injected photons is relatively small, they are redshifted and interact with intergalactic medium. 
Then, the intergalactic medium is partially ionized and the R-axion decay is regarded as an additional source of reionization. 
To be consistent with the observation of the optical depth to the last scattering surface, 
the R-axion abundance should be small enough. 
Assuming that the one-third of the energy of photons produced from R-axion decay that leaves the transparency window 
is converted to the ionization of the intergalactic medium, 
the R-axion abundance can be constrained from the inequality in Ref. \cite{Chen:2003gz,Zhang:2007zzh},
\begin{equation}
\log_{10} \zeta\lesssim \left\{ 
\begin{array}{ll}
6.77 + 3.96275x + 0.25858x^2 + 0.00445x^3,  & -17<x<-13 \\
-24.75-x,  & x< -17
\end{array}\right.
\end{equation}
where 
\begin{equation}
\zeta \equiv B_{\rm r} \rho_{a} /\rho_{\rm baryon}|_{\rm dec}=0.43\times 10^{10}{\rm GeV}^{-1}\left(\frac{\Omega_b h^2}{0.022}\right)^{-1} \frac{B_{\rm r} \rho_a}{s}, 
\end{equation}
and $x\equiv \log_{\rm 10} (\Gamma/{\rm s}^{-1})=-\log_{\rm 10} (\tau_a/{\rm s})$. 
Here $\Omega_b$  denotes the present density parameter of the baryonic matter. 
This constraint is complementary to that from the diffuse X-ray and $\gamma$-ray background. 

\subsubsection{Dark matter abundance}
If the lifetime of R-axions is longer than the present time $t_0$, 
most of R-axions remain the present Universe and contribute to the dark matter of the Universe. 
Thus, we can constrain the R-axion abundance in order not to exceed that of the dark matter. 
In terms of the energy-to-entropy ratio, the R-axion abundance is constrained as \cite{Komatsu:2010fb}
\begin{equation}
\frac{\rho_a}{s}< 4.7 \times 10^{-10} {\rm GeV} \left(\frac{\Omega_m h^2}{0.13} \right). \label{DMconst}
\end{equation}

\subsection{Constraints on model parameters}

\begin{figure}[htbp]
\begin{center}
	\hfill
  \includegraphics[width=.40\textwidth]{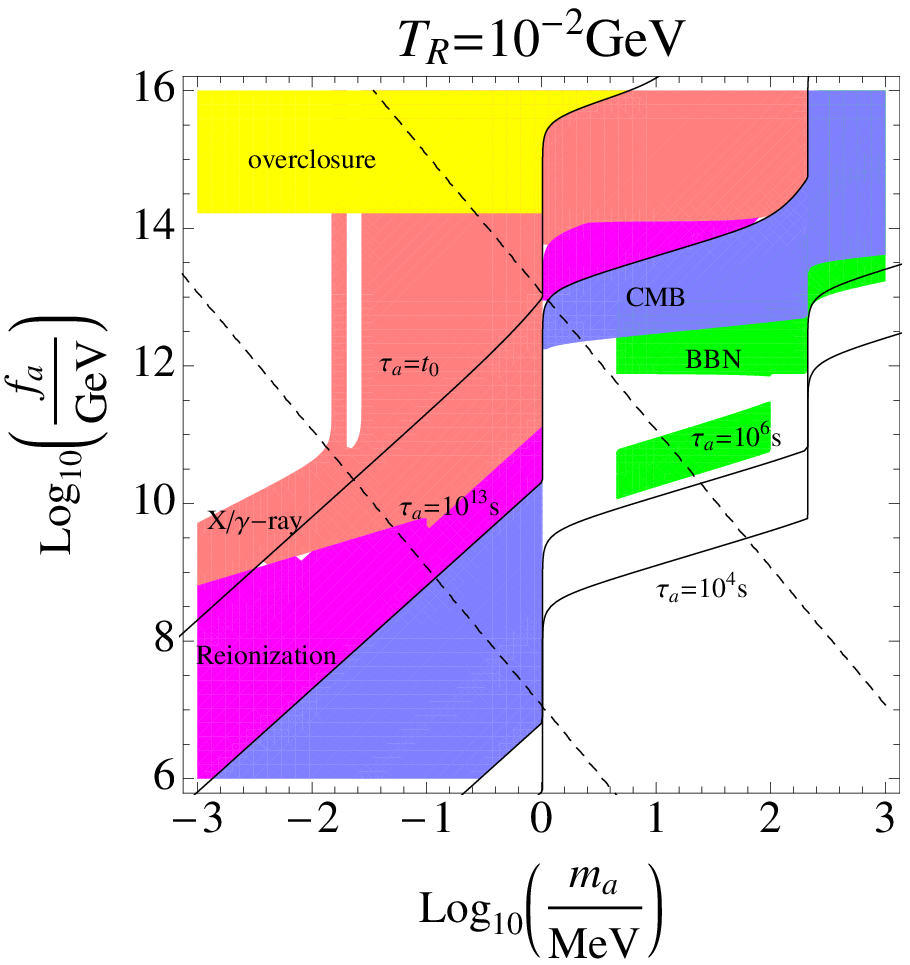}
	\hfill
  \includegraphics[width=.40\textwidth]{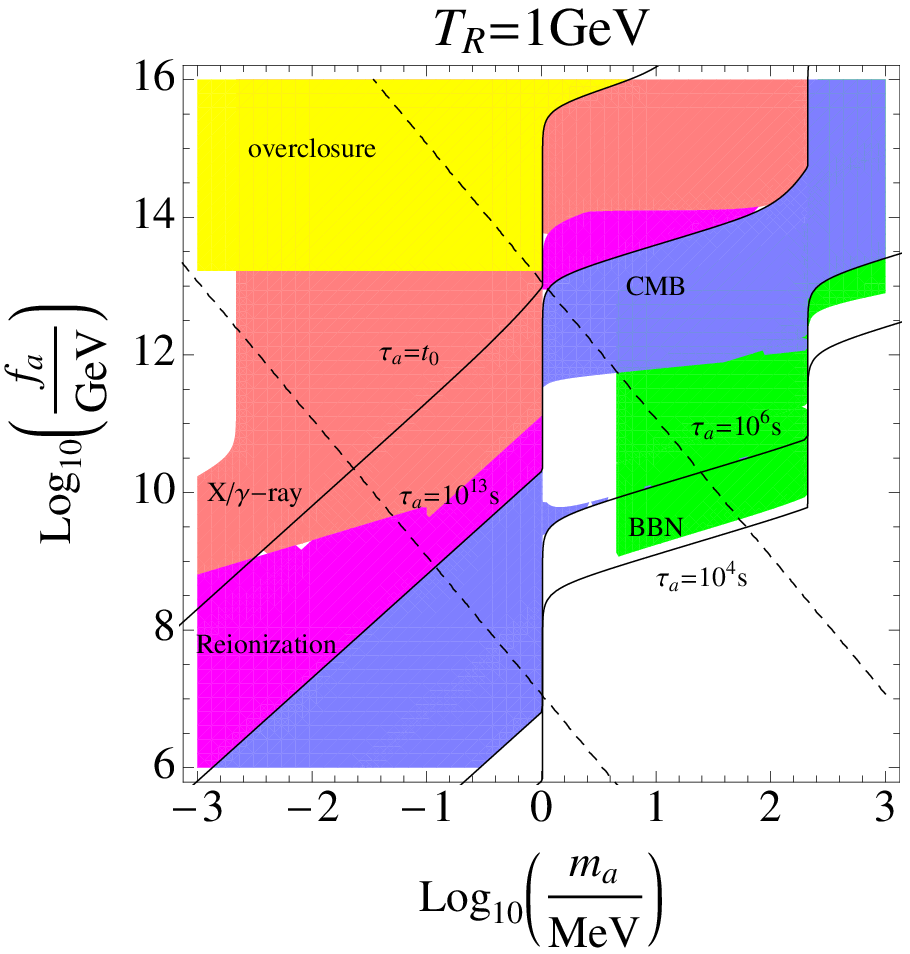}
	\hfill\mbox{}
\\
	\hfill
  \includegraphics[width=.40\textwidth]{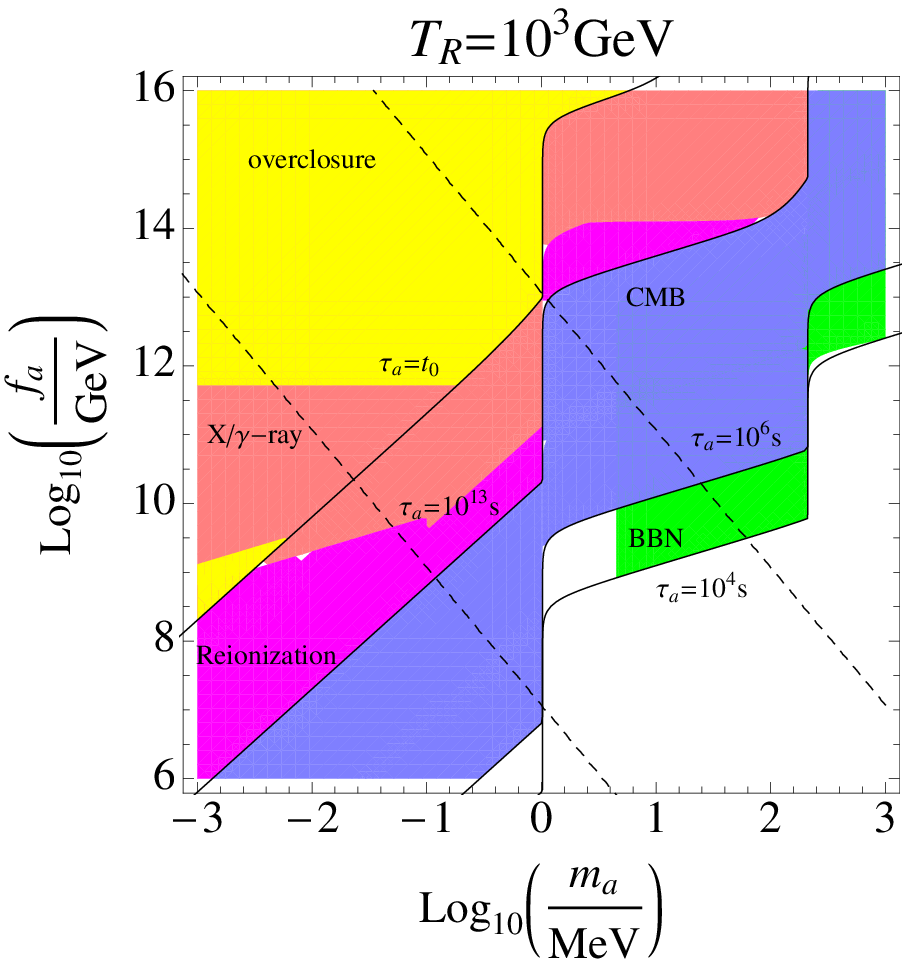}
	\hfill
  \includegraphics[width=.40\textwidth]{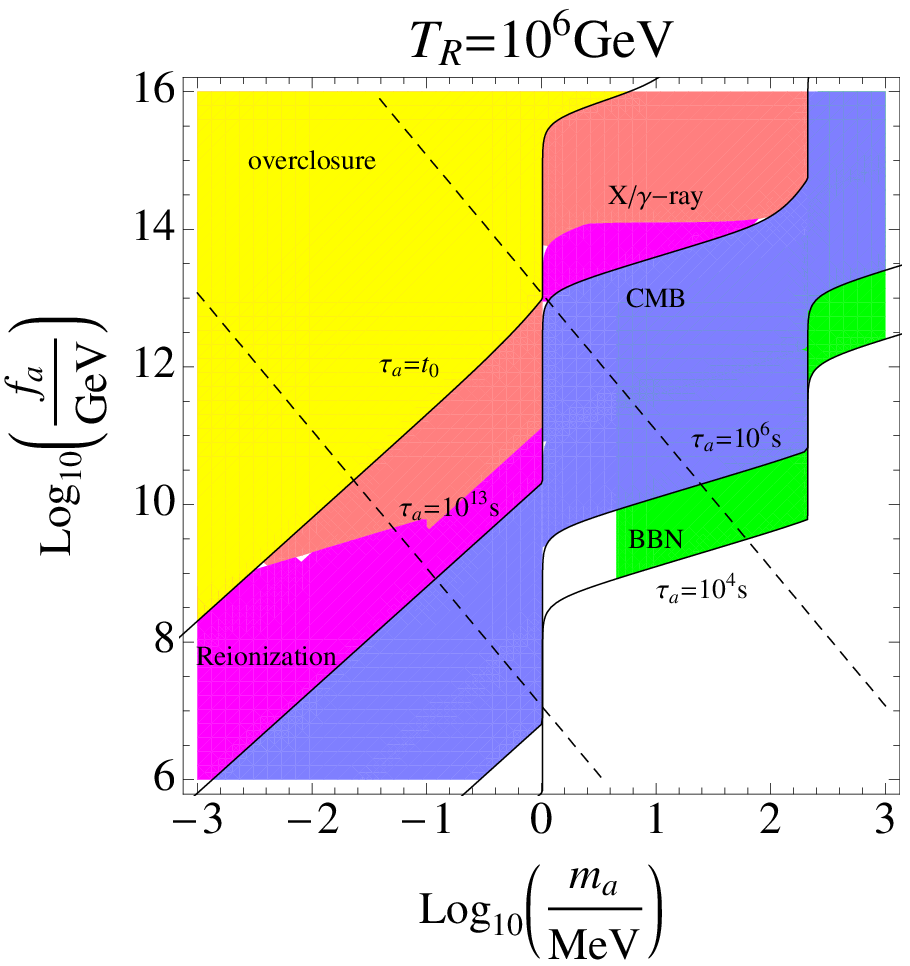}
	\hfill\mbox{}
  \caption{\sl Cosmological constraints on the model parameters, $m_a$ and $f_a$ with $T_R=10^{-2}$GeV, $1$GeV, $10^{3}$GeV, and $10^{6}$GeV. Each colored region is excluded and white region is allowed. Upper dotted lines and lower ones represent $m_{3/2}=1$keV and $m_{3/2}=1$eV. 
 Solid black lines represents the contour lines of equal R-axion lifetime, $\tau_a=10^4, 10^6, 10^{13}$ sec and $t_0$, 
  respectively.
}
\label{fig:constarint}
\end{center}
\end{figure}

Now we are ready to show cosmological constraints for spontaneous R-symmetry breaking models.  In Fig.\ref{fig:constarint}, we show the constraints on the model parameters, $m_a$ and $f_a$ coming from various conditions argued in the previous subsection. Each colored region is excluded and white region is allowed. As a reference, we also show lines of gravitino mass. Upper dotted lines and lower ones represent $m_{3/2}=1$keV, $m_{3/2}=1$eV, respectively. 
Here we focus on the region $f_a>10^6$ GeV since 
smaller $f_a$  is forbidden from laboratory experiments such as rare decays of $K^+$ or $B^0$ \cite{Andreas:2010ms}. 

For the higher reheating temperature, $T_R\gtrsim 10^2$ GeV, 
all the parameter space where R-axions decay at $t>10^6$ sec is ruled out regardless of reheating temperature, 
which comes from the CMB constraint. 
For $m_a<1$ MeV, it corresponds to $f_a\lesssim 10^7 {\rm GeV} (f_a/1{\rm MeV})^{3/2}$,
and $1 {\rm MeV} < m_a < 4.5$ MeV, it corresponds to $f_a<10^{9.5} {\rm GeV} (f_a/1{\rm MeV})^{1/2}$. 
For $m_a>4.5$ MeV, the BBN constraint opens and all the parameter space where the R-axion lifetime is $t>10^4$ sec 
is ruled out, again, regardless of reheating temeperature.   
For $4.5 {\rm MeV}<m_a<200$ MeV, it corresponds to $f_a\lesssim 10^9 {\rm GeV} (f_a/10 {\rm MeV})^{1/2}$, 
and for $m_a>200$ MeV, it corresponds to $f_a<10^{12} {\rm GeV}(f_a/200{\rm MeV})^{1/2}$. 
This is because R-axions are inevitably produced so much that cannot pass any constraints discussed above, 
especially, the BBN and CMB constraints. 
Short-lived R-axion is allowed because any entropy production before is not forbidden 
and there are no cosmological constraints. 

On the other hand, for the smaller reheating temperature, $T_R\lesssim 10^2$ GeV, 
several parameter space where R-axions decay later is allowed. 
This can be understood from Eqs.~\eqref{dynamics} and \eqref{th1}. 
For larger $f_a$, nonthermal production is dominant and the R-axion abundance 
is expressed as $\rho_a/s\propto f_a^{2}$, whereas thermal production, which 
depends on $f_a$ as $\rho_a \propto f_a^{-2}$, dominates for smaller $f_a$. 
Thus, the R-axion abundance takes its lower value at $f_a\sim 10^{11-12}$ GeV. 
As a result, allowed parameter region appears at $m_a\sim 10$ keV and 1-100 MeV for 
$f_a\sim 10^{12}$ GeV and $T_R\sim 10^{-2}-1$ GeV. 

One may regard that there is a parameter region where R-axions can be dark matter 
for smaller reheating temperature, $f_a\sim 10^{14}$ GeV. 
However, in this parameter region, R-saxion mass is considerably small, $m_s\lesssim 1$ MeV 
for a naive model discussed in Sec. \ref{sec2}. 
For the point of view of vacuum selection and R-string stability  \cite{OurpaperI}, 
this requires unacceptably small messenger mass. 
Therefore, we conclude that an ingenious model building is necessary for R-axions to be dark matter of the 
present Universe. 

Thus far, we did not take into account constraints from R-axinos or gravitinos. 
Gravitinos are produced from gluino scattering in thermal plasma and  their abundance is 
evaluated as \cite{Kawasaki:2004qu,Kawasaki:2004yh}, 
\begin{equation}
\frac{\rho_{3/2}}{s}\simeq 9.5 \times 10^{-8}{\rm GeV}\times \left(\frac{m_{\tilde g}}{1.5{\rm TeV}}\right)^2 \left(\frac{m_{3/2}}{15{\rm GeV}}\right)^{-1}\left(\frac{T_R}{10^{10}{\rm GeV}}\right), 
\end{equation}
where $m_{\tilde g}$ is the gaugino mass. 
Since gravitinos are stable in this case, gravitino abundance is constrained from dark matter abundance 
(Eq. \eqref{DMconst}). 
As a result, depending on the gaugino mass, another stringent constraint is imposed 
for model parameters in higher reheating temperature case $T_R\gtrsim 10$ GeV: 
Smaller $f_a$ and $m_a$ would be forbidden. 

In summary, we have shown that spontaneous R-symmetry breaking models are 
severely constrained from cosmological considerations and generally long-lived 
R-axions are forbidden. 
In order to avoid that, careful model building and smaller reheating temperature are required.

\section{Discussion \label{sec5}}

We have studied general cosmological constraints on spontaneous R-symmetry breaking models. 
We estimated the abundance of R-axion produced firstly via their dynamics such as coherent oscillation and 
decay of cosmic string/wall system, and secondly via thermal scattering process from gluon-axion interaction. 
It is interesting that R-axion production from R-string and wall systems are large enough and can be dominant 
in some parameter region. Basically the models were motivated by gauge mediation, gravitino as well as 
R-axion are relatively light. Therefore, R-axion tends to be long-lived. 
The conditions for the R-axion density coming from BBN, X-ray/$\gamma$-ray background, reionization and 
overclosure severely constrain the scale of R-symmetry breaking. 
As a result, smaller R-symmetry breaking scale and SUSY-breaking scale are disfavored from cosmological constraints. 
In the point of view of gauge mediation, this result weakens its motivation, but 
is consistent with the recent LHC results with 125 GeV Higgs-like boson and without SUSY particles \cite{SUSYiloHiggs}. 

It would be interesting to study further constraints for R-axion with relatively large mass. 
When the R-axion mass is larger than $1$ GeV, various decay channels to hadronic particle open. 
We expect that weaker but non-negligible constraints for large decay constant will be imposed, 
thought analysis would become involved.    

A phenomenologically viable model with long-lived R-axions can be constructed by introducing 
a mechanism 
diluting the R-axion density in the early universe. Although our primary interest was gauge mediation models, 
it would be easy to apply our analysis the closely related situations such as spontaneous R-symmetry breaking 
in thermal inflation models \cite{Thermal0}. 

In this paper, we have not explicitly shown a mechanism of vacuum selection of false vacuum. Existence of the R-string and walls are highly depend on the scenario of the early stage of universe. Also, as mentioned in the Introduction, imhomogenious vacuum decay by impurities such as a cosmic string depends on the details of the scenario \cite{OurpaperI}. So it may be useful to show an explicit example of full scenario and study R-axion cosmology in detail. This is beyond the scope of our study, so we will leave it as a future work.

\section*{Acknowledgment}

The authors would like to thank M. Eto for the early stage of collaboration, and 
 T.~Hiramatsu, Y.~Inoue, K.~Nakayama, A.~Ogasahara, A.~Ringwald, and T.~Takahashi  
for useful comments and discussions. 
KK would like to thank Kyoto University for their hospitality where this work was at the early stage. 
TK is supported in part by the Grant-in-Aid for the Global COE 
Program "The Next Generation of Physics, Spun from Universality and 
Emergence" and the JSPS Grant-in-Aid for Scientific Research
(A) No. 22244030 from the Ministry of Education, Culture,Sports, Science and 
Technology of Japan. YO's research is supported by The Hakubi Center for Advanced Research, Kyoto University.

\appendix

\section{Higgs sector}
\label{app:higgs-sector}

In this appendix, we show the Higgs sector and its mixing 
with the R-axion following Ref.~\cite{Goh:2008xz}.
Here, we concentrate on the neutral components, $H_u^0$ and $H^0_d$, of 
the Higgs sector in the minimal supersymmetric standard model.
The R-axion appears through the so-called B-term.
Then, the relevant terms in their scalar potential are given by 
\begin{eqnarray}
V&=& (|\mu|^2+m^2_{H_u})|H^0_u|^2 +  (|\mu|^2+m^2_{H_d})|H^0_d|^2 
+ \frac18 (g^2 + g'^2) ( |H^0_u|^2 - |H^0_d|^2 )^2 \nonumber \\
&& -(e^{i a/(\sqrt{2}f_a)}B\mu H_u^0 H_d^0 + c.c.),
\end{eqnarray}
where $\mu$ is the supersymmetric mass, originated from the 
$\mu$ term, $m^2_{H_u}$ and $m^2_{H_d}$ are soft SUSY breaking masses 
squared for $H_u$ and $H_d$, and $B\mu$ is the SUSY breaking B-term.
The B-term has an R-charge, and the axion appears there.

At the potential minimum, the Higgs fields develop their 
vacuum expectation values and the electroweak symmetry is broken.
Around the vacuum, we decompose the neutral Higgs fields as 
\begin{eqnarray}
H^0_u = \frac{1}{\sqrt{2}}(v_u + \rho_u)e^{i\xi_u/v_u}, 
\qquad 
H^0_d = \frac{1}{\sqrt{2}}(v_d + \rho_d)e^{i\xi_d/v_d}, 
\end{eqnarray}
where $v_u$ and $v_d$ are VEVs of $H^0_u$ and $H^0_d$.
We denote $v^2=  v_u^2 + v_d^2$, which is related to 
the $Z$-boson mass $m_Z$ as $v^2=4m_Z^2/(g^2 + g'^2) = (246)^2$ (GeV)$^2$.
Also we denote their ratio as 
\begin{eqnarray} 
\tan \beta = \frac{v_u}{v_d}.
\end{eqnarray}
Furthermore, the stationary conditions, 
\begin{eqnarray}
\frac{\partial V}{\partial H_u^0} = \frac{\partial V}{\partial H_d^0}
=0,
\end{eqnarray}
at $H_u^0=v_u$ and $H_d^0=v_d$, lead to the following relations:
\begin{eqnarray}
|\mu|^2+m^2_{H_u} = B\mu \cot \beta + \frac12 m_Z^2 \cos 2 \beta,
\nonumber \\
|\mu|^2+m^2_{H_d} = B\mu \tan \beta - \frac12 m_Z^2 \cos 2 \beta.
\end{eqnarray}
Using them, the mixing mass matrix of the axial parts, $\xi_{u,d}$ and 
the R-axion is given by 
\begin{eqnarray}
\frac{B\mu}{2}  (\xi_u, \xi_d,  a) \left(
\begin{array}{ccc}
\cot \beta & 1 & -r\cos \beta \\
1 & \tan \beta & - r \sin \beta \\
-r \cos \beta & -r \sin \beta & r^2 \sin \beta \cos \beta
\end{array}\right)
\left(
\begin{array}{c}
\xi_u \\ \xi_d \\  a
\end{array}
\right).
\end{eqnarray}
Then, the mass eigenstates are given by 
\begin{eqnarray}
\left(
\begin{array}{c}
G_0 \\ A_0 \\ \tilde a
\end{array}
\right) = \left(
\begin{array}{ccc}
\sin \beta & -\cot \beta & 0 \\
\kappa \cos \beta & \kappa \sin \beta 
& -\kappa r \sin \beta \cos \beta \\
\kappa r \cos^2 \beta \sin \beta 
& \kappa r \sin^2 \beta \cos \beta 
& \kappa 
\end{array}
\right)\left(
\begin{array}{c}
\xi_u \\ \xi_d \\  a
\end{array}
\right),
\end{eqnarray}
where $G_0$ and $\tilde a$ denote the would-be Nambu-Goldstone 
boson and low-energy R-axion, respectively.
 
\section{R-axion production from R-saxion decay \label{app2}}

Here we estimate the R-axion abundance from the R-saxion decay
and show that it is smaller than those from R-axion dynamics. 

Since we have assumed that the R-symmetry is restored in the early Universe, 
there should be homogeneous R-saxion oscillation associated with the 
spontaneous breaking of R-symmetry. 
It can take place when the Hubble parameter becomes smaller than the saxion mass. 
Here we assume that the R-symmetry is broken at $H=m_s$. 
If R-saxions receive thermal mass, the Hubble parameter at the time of phase transition 
becomes lower, but it requires large reheating temperature and we do not consider it here. 
The energy density of R-saxion is given by 
\begin{equation}
\rho_{s, {\rm osc}} (t_{\rm so}) \simeq m_s^2 f_a^2, 
\end{equation}
where the subscript ``so'' indicates that the parameter or variable is evaluated at the 
onset of R-saxion oscillation. 
The energy density of R-saxion oscillation decreases as $\rho_{s,{\rm osc}}\propto a^{-3}$ due to the Hubble expansion, 
and gradually R-saxion decays into R-axions at $H_{\rm sd}=\Gamma_{\rm sax}$. Here the subscript ``sd''
represents the parameter or variable is evaluated at R-saxion decay. 
The number density of R-axions from saxion decay, then, is evaluated as
\begin{equation}
n_{a,{\rm sax}} (t_{\rm sd})=\frac{2 \rho_{s, {\rm osc}}(t_{\rm sd})}{m_s}= \left\{
\begin{array}{ll}
\dfrac{2\Gamma_{\rm sax}^2}{m_s}f_a^2 & \text{for}, \quad H_{\rm sd}>H_R \\
2\dfrac{s_{\rm sd}}{s_R} \dfrac{H_R^2f_a^2}{m_s},  & \text{for} \quad H_{\rm so}>H_R>H_{\rm sd}  \\
2\dfrac{s_{\rm sd}}{s_{\rm so}} m_s f_a^2, & \text{for} \quad H_R>H_{\rm so}
\end{array}\right.
\end{equation}
where $s(T)=(2 \pi g_{*s}(T)/45)T^3$ is the entropy density. 
Such R-axions are relativistic at R-saxion decay and loose their energy due to the cosmic expansion. 
After some time, they become nonrelativistic. The energy-to-entropy ratio, $\rho_{a, {\rm sax}}/s$, 
is fixed at that time (if reheating is completed.) 
Therefore, we can estimate the abundance of R-axions from R-saxion decay as
\begin{equation}
\frac{\rho_{a,{\rm sax}}}{s}=\frac{m_a n_{a, {\rm sax}}}{s}= \left\{
\begin{array}{ll}
\dfrac{m_a f_a^2}{2 m_s M_{\rm pl}^2}T_R, &\text{for} \quad H_{\rm so}>H_R \\
\dfrac{45}{\pi^2g_{*s}(T_{\rm so})} \dfrac{m_a m_s f_a^2}{T_{\rm so}^3}, &\text{for} \quad H_{\rm so}<H_R
\end{array}\right. 
\end{equation}
with $T_{\rm so}=(\pi^2 g_*(T_{\rm so})/90)^{1/4} (m_sM_{\rm pl})^{1/2}$. 
Here we neglected the interaction of R-axion and assumed that the number of R-axions in a comoving 
volume is conserved. 
We can easily show that the abundance of R-axions from R-saxion decay is always smaller than that from 
R-string and R-string-wall system. 
Note that if the phase transition is driven by thermal potential, R-axion abundance from R-saxion decay 
is larger than that is estimated above. 
However, as noted, it requires high reheating temperature, in which the model 
has already been constrained by the thermal contribution strictly. 
As a result, the conclusion does not change.



\begin{thebibliography}{1}
 
\bibitem{LHChiggs} 
  G.~Aad {\it et al.}  [ATLAS Collaboration],
  Phys.\ Lett.\ B {\bf 716}, 1 (2012)
  [arXiv:1207.7214 [hep-ex]]; 
  S.~Chatrchyan {\it et al.}  [CMS Collaboration],
  Phys.\ Lett.\ B {\bf 716}, 30 (2012)
  [arXiv:1207.7235 [hep-ex]].
  
\bibitem{SUSYiloHiggs} 
  P.~Draper, P.~Meade, M.~Reece and D.~Shih,
  Phys.\ Rev.\ D {\bf 85}, 095007 (2012)
  [arXiv:1112.3068 [hep-ph]]; 
  B.~Bhattacherjee, B.~Feldstein, M.~Ibe, S.~Matsumoto and T.~T.~Yanagida,
  Phys.\ Rev.\ D {\bf 87}, 015028 (2013)
  [arXiv:1207.5453 [hep-ph]]; 
  T.~Higaki, K.~Kamada and F.~Takahashi,
  JHEP {\bf 1209}, 043 (2012)
  [arXiv:1207.2771 [hep-ph]];
  B.~Feldstein and T.~T.~Yanagida,
  Phys.\ Lett.\ B {\bf 720}, 166 (2013)
  [arXiv:1210.7578 [hep-ph]]; 
  T.~Moroi, T.~T.~Yanagida and N.~Yokozaki,
  Phys.\ Lett.\ B {\bf 719}, 148 (2013)
  [arXiv:1211.4676 [hep-ph]]; 
  M.~Endo, K.~Hamaguchi, S.~Iwamoto and N.~Yokozaki,
  JHEP {\bf 1206}, 060 (2012)
  [arXiv:1202.2751 [hep-ph]].
  



\bibitem{rev1} 
  K.~A.~Intriligator and N.~Seiberg,
  Class.\ Quant.\ Grav.\  {\bf 24}, S741 (2007)
  [hep-ph/0702069].


\bibitem{rev2} 
  R.~Kitano, H.~Ooguri and Y.~Ookouchi,
  Ann.\ Rev.\ Nucl.\ Part.\ Sci.\  {\bf 60}, 491 (2010)
  [arXiv:1001.4535 [hep-th]].

\bibitem{rev3} 
  M.~Dine and J.~D.~Mason,
  Rept.\ Prog.\ Phys.\  {\bf 74}, 056201 (2011)
  [arXiv:1012.2836 [hep-th]].






\bibitem{Kitano1} 
  R.~Kitano,
  Phys.\ Lett.\ B {\bf 641}, 203 (2006)
  [hep-ph/0607090].
  
\bibitem{Kitano2} 
  M.~Ibe and R.~Kitano,
  JHEP {\bf 0708}, 016 (2007)
  [arXiv:0705.3686 [hep-ph]].

\bibitem{Intriligator:2007py} 
 K.~A.~Intriligator, N.~Seiberg and D.~Shih,
JHEP {\bf 0707}, 017 (2007)  [hep-th/0703281].  

\bibitem{KOO} 
  R.~Kitano, H.~Ooguri and Y.~Ookouchi,
  Phys.\ Rev.\ D {\bf 75}, 045022 (2007)
  [hep-ph/0612139].

\bibitem{Abe:2007ax} 
 H.~Abe, T.~Kobayashi and Y.~Omura,
JHEP {\bf 0711}, 044 (2007)  [arXiv:0708.3148 [hep-th]].  

\bibitem{R1} 
  J.~L.~Evans, M.~Ibe, M.~Sudano and T.~T.~Yanagida,
  JHEP {\bf 1203}, 004 (2012)
  [arXiv:1103.4549 [hep-ph]].

\bibitem{R2} 
  J.~Goodman, M.~Ibe, Y.~Shirman and F.~Yu,
  Phys.\ Rev.\ D {\bf 84}, 045015 (2011)
  [arXiv:1106.1168 [hep-th]].
  
\bibitem{Kang:2012fn} 
  Z.~Kang, T.~Li and Z.~Sun,
  arXiv:1209.1059 [hep-th].

\bibitem{Ferretti:2007ec} 
 L.~Ferretti,
JHEP {\bf 0712}, 064 (2007)  [arXiv:0705.1959 [hep-th]].  

\bibitem{Cho:2007yn} 
 H.~Y.~Cho and J.~-C.~Park,
JHEP {\bf 0709}, 122 (2007)  [arXiv:0707.0716 [hep-ph]].  

\bibitem{Abel:2007jx} 
 S.~Abel, C.~Durnford, J.~Jaeckel and V.~V.~Khoze,
Phys.\ Lett.\ B {\bf 661}, 201 (2008)  [arXiv:0707.2958 [hep-ph]].  

\bibitem{Aldrovandi:2008sc} 
 L.~G.~Aldrovandi and D.~Marques,
JHEP {\bf 0805}, 022 (2008)  [arXiv:0803.4163 [hep-th]].  

\bibitem{Carpenter:2008wi} 
 L.~M.~Carpenter, M.~Dine, G.~Festuccia and J.~D.~Mason,
Phys.\ Rev.\ D {\bf 79}, 035002 (2009)  [arXiv:0805.2944 [hep-ph]].  

\bibitem{Giveon:2008ne} 
 A.~Giveon, A.~Katz, Z.~Komargodski and D.~Shih,
JHEP {\bf 0810}, 092 (2008)  [arXiv:0808.2901 [hep-th]].  

\bibitem{Higaki:2011bz} 
  T.~Higaki and R.~Kitano,
  Phys.\ Rev.\ D {\bf 86}, 075027 (2012)
  [arXiv:1104.0170 [hep-ph]].


\bibitem{Nelson:1993nf} 
 A.~E.~Nelson and N.~Seiberg,
Nucl.\ Phys.\ B {\bf 416}, 46 (1994)  [hep-ph/9309299].  



\bibitem{Shih:2007av}
 D.~Shih,
 JHEP {\bf 0802}, 091 (2008)
 [hep-th/0703196].

\bibitem{Extra} 
  C.~Cheung, A.~L.~Fitzpatrick and D.~Shih,
  JHEP {\bf 0807}, 054 (2008)
  [arXiv:0710.3585 [hep-ph]].
  
\bibitem{KS} 
 Z.~Komargodski and D.~Shih,
JHEP {\bf 0904}, 093 (2009)  [arXiv:0902.0030 [hep-th]].  

\bibitem{Kibble} 
  T.~W.~B.~Kibble,
 J.\ Phys.\ A {\bf 9}, 1387 (1976).  

\bibitem{Zurek}
W. H. Zurek,
Nature {\bf 317}, 505-508 (1985);
Phys. Rep. {\bf 276}, 177-221 (1996).

\bibitem{Pole} 
 P.~J.~Steinhardt,
Nucl.\ Phys.\ B {\bf 190}, 583 (1981); 

\bibitem{Hosotani} 
  Y.~Hosotani,
  Phys.\ Rev.\ D {\bf 27}, 789 (1983).

\bibitem{OurpaperI}
  M.~Eto, Y.~Hamada, K.~Kamada, T.~Kobayashi, K.~Ohashi and Y.~Ookouchi,
  JHEP {\bf 1303}, 159 (2013)
  [arXiv:1211.7237 [hep-th]].
  
\bibitem{NakaiOokouchi} 
  Y.~Nakai and Y.~Ookouchi,
  JHEP {\bf 1101}, 093 (2011)
  [arXiv:1010.5540 [hep-th]].

\bibitem{AzeyanagiKobayashi}
  T.~Azeyanagi, T.~Kobayashi, A.~Ogasahara and K.~Yoshioka,
  JHEP {\bf 1109}, 112 (2011)
  [arXiv:1106.2956 [hep-ph]]; 
  Phys.\ Rev.\ D {\bf 86}, 095026 (2012)
  [arXiv:1208.0796 [hep-ph]].


\bibitem{Goh:2008xz} 
  H.~-S.~Goh and M.~Ibe,
  JHEP {\bf 0903}, 049 (2009)
  [arXiv:0810.5773 [hep-ph]].
  
\bibitem{Sikivie:1982qv} 
  P.~Sikivie,
  Phys.\ Rev.\ Lett.\  {\bf 48}, 1156 (1982);
  D.~H.~Lyth,
  Phys.\ Lett.\ B {\bf 275}, 279 (1992);
  M.~Nagasawa and M.~Kawasaki,
  Phys.\ Rev.\ D {\bf 50}, 4821 (1994)
  [astro-ph/9402066];
  S.~Chang, C.~Hagmann and P.~Sikivie,
  Phys.\ Rev.\ D {\bf 59}, 023505 (1999)
  [hep-ph/9807374];
\bibitem{Hiramatsu:2012gg} 
  T.~Hiramatsu, M.~Kawasaki, K.~'i.~Saikawa and T.~Sekiguchi,
  Phys.\ Rev.\ D {\bf 85}, 105020 (2012)
  [Erratum-ibid.\ D {\bf 86}, 089902 (2012)]
  [arXiv:1202.5851 [hep-ph]].
  
\bibitem{Dine:1982ah} 
  J.~Preskill, M.~B.~Wise and F.~Wilczek,
  Phys.\ Lett.\ B {\bf 120}, 127 (1983);
  L.~F.~Abbott and P.~Sikivie,
  Phys.\ Lett.\ B {\bf 120}, 133 (1983);
  M.~Dine and W.~Fischler,
  Phys.\ Lett.\ B {\bf 120}, 137 (1983); 

  
\bibitem{Davis:1986xc} 
  R.~L.~Davis,
  Phys.\ Lett.\ B {\bf 180}, 225 (1986); 
  A.~Vilenkin and T.~Vachaspati,
  Phys.\ Rev.\ D {\bf 35}, 1138 (1987);
  R.~A.~Battye and E.~P.~S.~Shellard,
  Nucl.\ Phys.\ B {\bf 423}, 260 (1994)
  [astro-ph/9311017].
  
\bibitem{Yamaguchi:1998gx} 
  M.~Yamaguchi, M.~Kawasaki and J.~'i.~Yokoyama,
  Phys.\ Rev.\ Lett.\  {\bf 82}, 4578 (1999)
  [hep-ph/9811311].
  
\bibitem{Hiramatsu:2010yu} 
  T.~Hiramatsu, M.~Kawasaki, T.~Sekiguchi, M.~Yamaguchi and J.~'i.~Yokoyama,
  Phys.\ Rev.\ D {\bf 83}, 123531 (2011)
  [arXiv:1012.5502 [hep-ph]].
  
\bibitem{Yamaguchi:2002zv} 
  M.~Yamaguchi and J.~'i.~Yokoyama,
  Phys.\ Rev.\ D {\bf 66}, 121303 (2002)
  [hep-ph/0205308]; 
  M.~Yamaguchi and J.~'i.~Yokoyama,
  Phys.\ Rev.\ D {\bf 67}, 103514 (2003)
  [hep-ph/0210343].
  
  
  \bibitem{stringreveiw}
  A.~Vilenkin and E.~P.~S.~Shellard, {\it Cosmic Strings and Other Topological Defects}, 
  Cambridge University Press, Cambridge, England, 2000, 
  M.~B.~Hindmarsh and T.~W.~B.~Kibble; 
  {\it Cosmic strings}
  Rept.\ Prog.\ Phys.\  {\bf 58}, 477 (1995)
  [arXiv:hep-ph/9411342].

  
  
\bibitem{Masso:2002np} 
 E.~Masso, F.~Rota and G.~Zsembinszki,
 Phys.\ Rev.\ D {\bf 66}, 023004 (2002)  [hep-ph/0203221].  
 
\bibitem{Sikivie:2006ni} 
  P.~Sikivie,
  Lect.\ Notes Phys.\  {\bf 741}, 19 (2008)
  [astro-ph/0610440].
  
\bibitem{Graf:2010tv} 
  P.~Graf and F.~D.~Steffen,
  Phys.\ Rev.\ D {\bf 83}, 075011 (2011)
  [arXiv:1008.4528 [hep-ph]].
  
  
  
\bibitem{Kawasaki:2007mk} 
  M.~Kawasaki, K.~Nakayama and M.~Senami,
  JCAP {\bf 0803}, 009 (2008)
  [arXiv:0711.3083 [hep-ph]].
  
\bibitem{Kawasaki:2004qu} 
  M.~Kawasaki, K.~Kohri and T.~Moroi,
  Phys.\ Rev.\ D {\bf 71}, 083502 (2005)
  [astro-ph/0408426].
  
\bibitem{Fixsen:1996nj} 
  D.~J.~Fixsen, E.~S.~Cheng, J.~M.~Gales, J.~C.~Mather, R.~A.~Shafer and E.~L.~Wright,
  Astrophys.\ J.\  {\bf 473}, 576 (1996)
  [astro-ph/9605054]; 
  K.~Hagiwara {\it et al.}  [Particle Data Group Collaboration],
  Phys.\ Rev.\ D {\bf 66}, 010001 (2002).
  
\bibitem{Ellis:1990nb} 
  J.~R.~Ellis, G.~B.~Gelmini, J.~L.~Lopez, D.~V.~Nanopoulos and S.~Sarkar,
  Nucl.\ Phys.\ B {\bf 373}, 399 (1992).
  
\bibitem{Hu:1993gc} 
  W.~Hu and J.~Silk,
  Phys.\ Rev.\ Lett.\  {\bf 70}, 2661 (1993).
  
  
  
\bibitem{Chen:2003gz} 
  X.~-L.~Chen and M.~Kamionkowski,
  Phys.\ Rev.\ D {\bf 70}, 043502 (2004)
  [astro-ph/0310473]; 
  T.~R.~Slatyer, N.~Padmanabhan and D.~P.~Finkbeiner,
  Phys.\ Rev.\ D {\bf 80}, 043526 (2009)
  [arXiv:0906.1197 [astro-ph.CO]].
  
\bibitem{ASCA} Gendreau, K.~C., 
Mushotzky, R., Fabian, A.~C., et al. 1995 \ Publ.\ Astron.\ Soc.\ Japan {\bf 47} L5. 

\bibitem{HEAO} Gruber, D.~E., Matteson, 
J.~L., Peterson, L.~E., \& Jung, G.~V.\ Astrophys.\ J.\  {\bf 520}, 124 (1999). 
  
\bibitem{1998PhDT3K} Kappadath, S.~C.\ 1998, 
Ph.D.~Thesis,  

\bibitem{Sreekumar:1997un} 
  P.~Sreekumar {\it et al.}  [EGRET Collaboration],
  Astrophys.\ J.\  {\bf 494}, 523 (1998)
  [astro-ph/9709257].

\bibitem{Abdo:2010nz} 
  A.~A.~Abdo {\it et al.}  [Fermi-LAT Collaboration],
  Phys.\ Rev.\ Lett.\  {\bf 104}, 101101 (2010)
  [arXiv:1002.3603 [astro-ph.HE]].
  

\bibitem{Worsley:2004tc} 
  M.~A.~Worsley, A.~C.~Fabian, F.~E.~Bauer, D.~M.~Alexander, G.~Hasinger, S.~Mateos, H.~Brunner and W.~N.~Brandt {\it et al.},
  Mon.\ Not.\ Roy.\ Astron.\ Soc.\  {\bf 357}, 1281 (2005)
  [astro-ph/0412266].

\bibitem{Hickox:2005dz} 
  R.~C.~Hickox and M.~Markevitch,
  Astrophys.\ J.\  {\bf 645}, 95 (2006)
  [astro-ph/0512542].

\bibitem{CXBfit} Ajello, M., Greiner, J., 
Sato, G., et al.\  Astrophys.\ J.\  {\bf 689}, 666 (2008)






  
\bibitem{Zhang:2007zzh} 
  L.~Zhang, X.~Chen, M.~Kamionkowski, Z.~-g.~Si and Z.~Zheng,
  Phys.\ Rev.\ D {\bf 76}, 061301 (2007)
  [arXiv:0704.2444 [astro-ph]].
  
  
\bibitem{Komatsu:2010fb} 
  E.~Komatsu {\it et al.}  [WMAP Collaboration],
  Astrophys.\ J.\ Suppl.\  {\bf 192}, 18 (2011)
  [arXiv:1001.4538 [astro-ph.CO]].
  
\bibitem{Andreas:2010ms} 
  S.~Andreas, O.~Lebedev, S.~Ramos-Sanchez and A.~Ringwald,
  JHEP {\bf 1008}, 003 (2010)
  [arXiv:1005.3978 [hep-ph]].

  
\bibitem{Kawasaki:2004yh}
  M.~Kawasaki, K.~Kohri and T.~Moroi,
  Phys.\ Lett.\  B {\bf 625}, 7 (2005)
  [arXiv:astro-ph/0402490]. 

\bibitem{Thermal0} 
  D.~H.~Lyth and E.~D.~Stewart,
  Phys.\ Rev.\ D {\bf 53}, 1784 (1996)
  [hep-ph/9510204]; 
  T.~Asaka, J.~Hashiba, M.~Kawasaki and T.~Yanagida,
  Phys.\ Rev.\ D {\bf 58}, 083509 (1998)
  [hep-ph/9711501]; T.~Moroi and K.~Nakayama,
  Phys.\ Lett.\ B {\bf 703}, 160 (2011)
  [arXiv:1105.6216 [hep-ph]].
  
\end{thebibliography}
\end{document}